\documentclass[prb, reprint, superscriptaddress,floatfix]{revtex4-2}

\usepackage{amsmath}
\usepackage{graphicx}
\usepackage{verbatim}
\usepackage{xcolor}
\usepackage{hyperref}
\usepackage{url}

\begin{document}
\title{Transport signatures of Fermi arcs at twin boundaries in Weyl materials}

\author{Sahal Kaushik}
\email{sahal.kaushik@su.se}
\affiliation{Department of Physics and Astronomy, Stony Brook University, Stony Brook, NY 11794, USA}
\affiliation{Nordita, Stockholm University and KTH Royal Institute of Technology, Hannes Alfv\'{e}ns v\"{a}g 12, SE-106 91 Stockholm, Sweden}

\author{I\~{n}igo Robredo}
\affiliation{Max Planck Institute for Chemical Physics of Solids, Dresden D-01187, Germany}
\affiliation{Donostia International Physics Center, 20018 Donostia-San Sebasti\'{a}n, Spain}

\author{Nitish Mathur}
\affiliation{Department of Chemistry, University of Wisconsin—Madison, 1101 University Avenue, Madison, Wisconsin 53706, USA}
\affiliation{Department of Chemistry, Princeton University, Princeton, New Jersey 08544, USA}

\author{Leslie M. Schoop}
\affiliation{Department of Chemistry, Princeton University, Princeton, New Jersey 08544, USA}

\author{Song Jin}
\affiliation{Department of Chemistry, University of Wisconsin—Madison, 1101 University Avenue, Madison, Wisconsin 53706, USA}

\author{Maia G. Vergniory}
 \affiliation{Max Planck Institute for Chemical Physics of Solids, Dresden D-01187, Germany}
\affiliation{Donostia International Physics Center, 20018 Donostia-San Sebasti\'{a}n, Spain}

\author{Jennifer Cano}
\email{jennifer.cano@stonybrook.edu}
\affiliation{Department of Physics and Astronomy, Stony Brook University, Stony Brook, NY 11794, USA}
\affiliation{Center for Computational Quantum Physics, Flatiron Institute, New York, NY 10010, USA}

\begin{abstract}
One of the most striking signatures of Weyl fermions is their surface Fermi arcs. Less known is that Fermi arcs can also be localized at  internal twin boundaries where two Weyl materials of opposite chirality meet. In this work, we derive constraints on the topology and connectivity of these ``internal Fermi arcs.'' We show that internal Fermi arcs can exhibit transport signatures and propose two probes: quantum oscillations and a quantized chiral magnetic current. We propose merohedrally twinned B20 materials as candidates to host internal Fermi arcs, verified through both model and ab initio calculations. Our theoretical investigation sheds lights on the topological features and motivates experimental studies into the intriguing physics of internal Fermi arcs.
\end{abstract}
\maketitle

\section{Introduction}
Weyl fermions are topologically protected point degeneracies in the energy spectrum of three-dimensional crystals. 
Their topology leads to interesting properties:
since Weyl points are quantized monopole charges of Berry curvature, they contribute to unusual bulk transport properties, including an anomaly-induced \cite{AnomalyAdler,AnomalyBJ} negative quadratic longitudinal magnetoresistance \cite{CME,son2013chiral,burkov2014chiral,nielsen1983adler,li2016chiral,Xiong2015,Zhang2016,Huang2015} and a non-saturating Nernst effect \cite{SkinnerNernst}. In addition, they exhibit nontrivial optical responses, including circular photocurrents \cite{PLee,MaTaAs} that can be quantized \cite{deJuanQCPE}.

But perhaps the most striking signature of the topology of Weyl fermions is their surface Fermi arcs, 
i.e., gapless states that connect the surface projections of Weyl cones of opposite chirality \cite{Wan11}. 
Detecting Fermi arcs via ARPES is the canonical experimental signature of Weyl cones \cite{TaAsArcs,NbAsArcs,schroter2019chiral,schroter2020observation, CoS2_exp}.
Fermi arcs can also be detected via quantum oscillations \cite{ArcOscillations1,ArcOscillations2,ArcOscillations3,ArcOscillationsExpt} and photocurrents \cite{ArcCurrent1,ArcCurrent2,ArcCurrent3}.

The topological arguments for the existence of Fermi arcs at external surfaces also apply to twin boundaries inside of the crystal \cite{schroter2020observation}.
Specifically, a twinned crystal contains a defect plane that separates domains in which the crystal has grown in different orientations. 
If the two domains are related by an orientation-reversing symmetry (which flips the chirality of bulk Weyl points),
then there must be Fermi arcs localized at the domain wall, as we will shortly derive.
Similar Fermi arcs can appear at twisted interfaces \cite{TwistArcs1,TwistArcs2} and domain walls in magnetic Weyl materials \cite{MagWallArcs1,MagWallArcs2,MagWallArcs3}.
We refer to Fermi arcs localized inside the crystal at the twin boundary as ``internal'' Fermi arcs, as opposed to the more familiar ``external'' Fermi arcs, which exist at the boundary with the vacuum.

The possibility of internal Fermi arcs was proposed in Ref \cite{schroter2020observation}.
However, because the internal Fermi arcs are surrounded by conducting and opaque bulk crystals on both sides of the boundary, they are not easily accessible by ARPES and photocurrent measurements.
Thus, how to observe them has been an open problem.

We propose to solve this problem by describing two transport phenomena -- quantum oscillations and a quantized chiral magnetic effect -- that can detect internal Fermi arcs.
We first demonstrate these effects in a minimal model of four Weyl cones and
then apply our results to the B20 materials where internal Fermi arcs were first proposed \cite{schroter2020observation}. We calculate the bulk spectrum and internal Fermi arcs for a simplified tight-binding model and demonstrate that the connectivity of the arcs depends on the microsopic details of the interface and can change with energy. 
We also show that the localization of the ends of the Fermi arcs to the domain wall can be controlled by an external magnetic field.
We then compute the bulk spectrum and internal arcs for a twinned slab of the B20 material CoSi in an ab initio calculation and show that it agrees qualitatively with our model.
Thus, CoSi is a promising candidate to observe the predicted transport phenomena.

\section{Chirality of Weyl Fermions}

The simplest Weyl Hamiltonian is an isotropic linear two band model:
\begin{equation}
    H = \chi v_0\vec{k}\cdot\vec{\sigma},
    \label{eq:H0}
\end{equation}
where $\chi = \pm 1$ is the chirality, which is $+1$ for right handed and $-1$ for  left handed fermions. The Berry curvature of the eigenstates is:
\begin{equation}
    \vec{\Omega} = \pm \chi \frac{1}{2k^2}\hat{k},
\end{equation}
where $+$ is for the upper band and $-$ for the lower band.

The flux of Berry curvature through a surface enclosing a Weyl point is always quantized in units of $2\pi$: $\oint\vec{\Omega}\cdot d\hat{n} = 2\pi C$. The integer $C$ is called the Chern number and is equal to $\chi = \pm 1$ for the Hamiltonian in Eq.~(\ref{eq:H0}).
Since the total Chern number inside the entire Brillouin zone must vanish, Weyl fermions always come in pairs of opposite chirality.

More generally, a Weyl fermion can have anisotropy:
\begin{equation}
    H =  v^i_a k_i \sigma_a + v^i_t k_i.
\end{equation}
The chirality is given by $\chi = \mathrm{sgn}\,\mathrm{det}(v^i_a)$.

In certain space groups, symmetry can protect twofold quadratic \cite{SrSi2} and cubic \cite{CubicWeyl,FourWeyl1,FourWeyl2} fermions and multifold chiral fermions (including double spin-1/2, spin-1, and  spin-3/2 fermions) \cite{CanoMultifold,cano2019multifold}. These fermions  generally have Chern numbers greater than $\pm 1$. If the symmetries protecting them are broken, they split into multiple Weyl cones with Chern number $\pm 1$.


\section{Topology and Connectivity of Fermi Arcs}
\label{sec:Fermiarcs}


The topology of the chiral bulk states results in Fermi arcs connecting the projections of Weyl points at surfaces \cite{Wan11}, as we now review.  
Consider a cylinder in the bulk Brillouin zone. Its projection onto the surface Brillouin zone is a closed loop. If the cylinder has a non-zero Chern number, the corresponding loop exhibits gapless surface states, whose number is equal to the Chern number. Since a surface enclosing a Weyl cone has a non-zero Berry flux, a loop in the surface Brillouin zone surrounding the projection of a Weyl cone has a gapless surface state. The set of all such states coming from different cylinders enclosing the Weyl cone together comprise a Fermi arc. Thus, the number of Fermi arcs emanating from the surface projection of each Weyl point is equal to its Chern number.  If the surface projections of two or more Weyl points coincide, the number of Fermi arcs emanating from their shared projection onto the surface Brillouin zone is the sum of their Chern numbers, i.e., if two simple Weyl cones have the same chirality, there are two arcs, but if they have opposite chirality, there are no arcs.  Each Fermi arc starts and ends at the projections of bulk Weyl points with opposite chirality.



\begin{figure}
    \centering
    \includegraphics[scale=0.2]{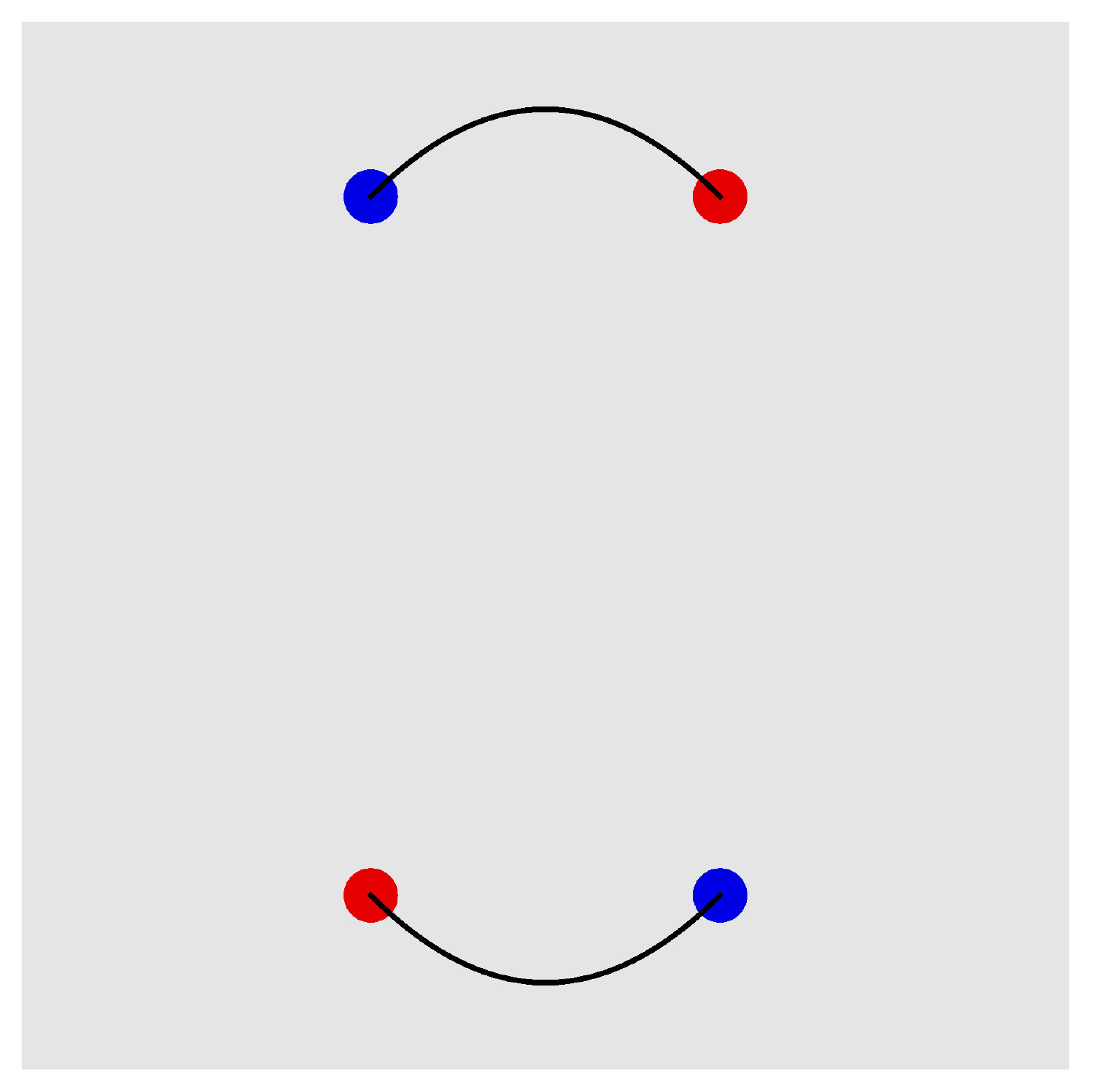}\includegraphics[scale=0.2]{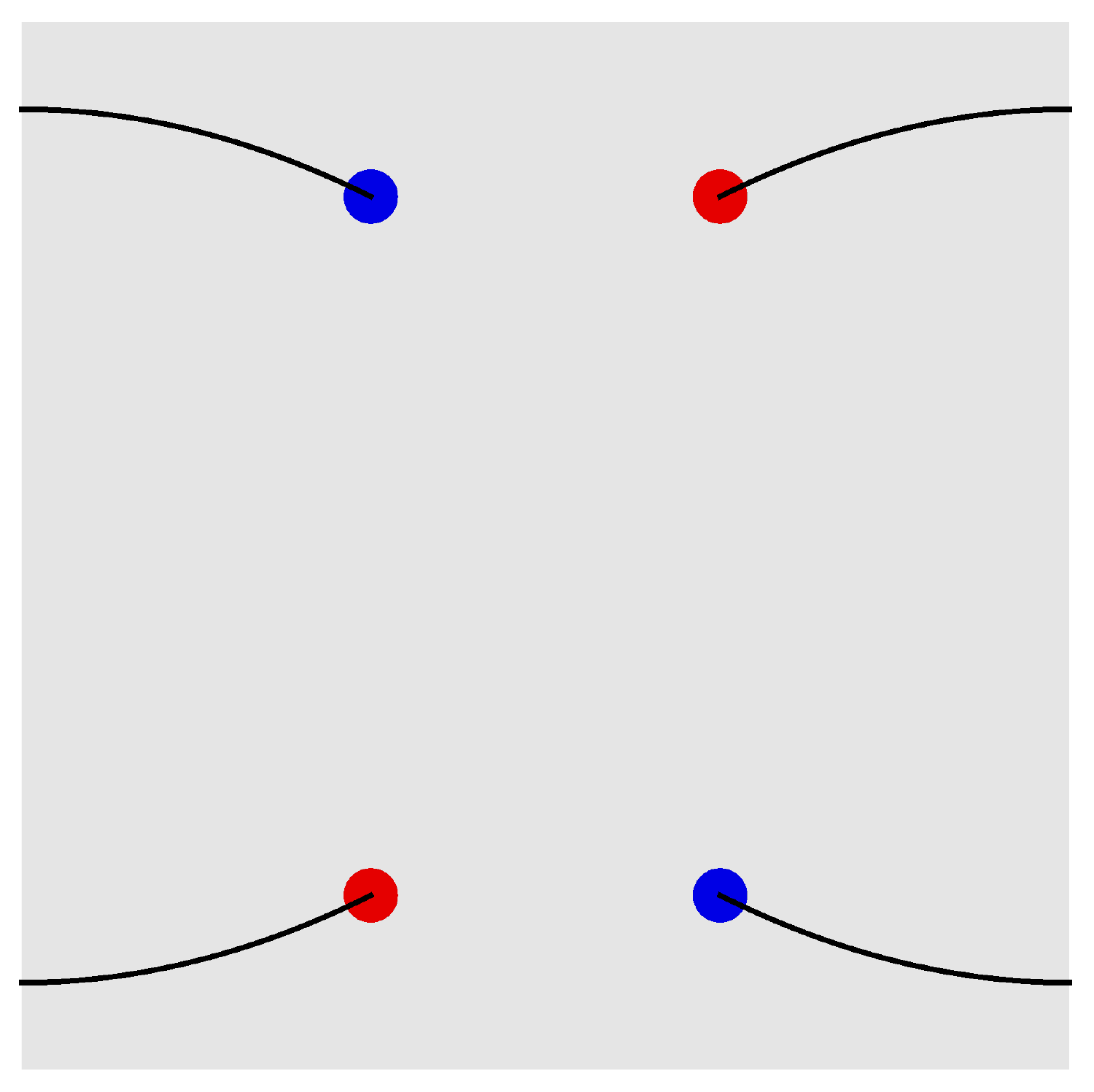}\\ \includegraphics[scale=0.2]{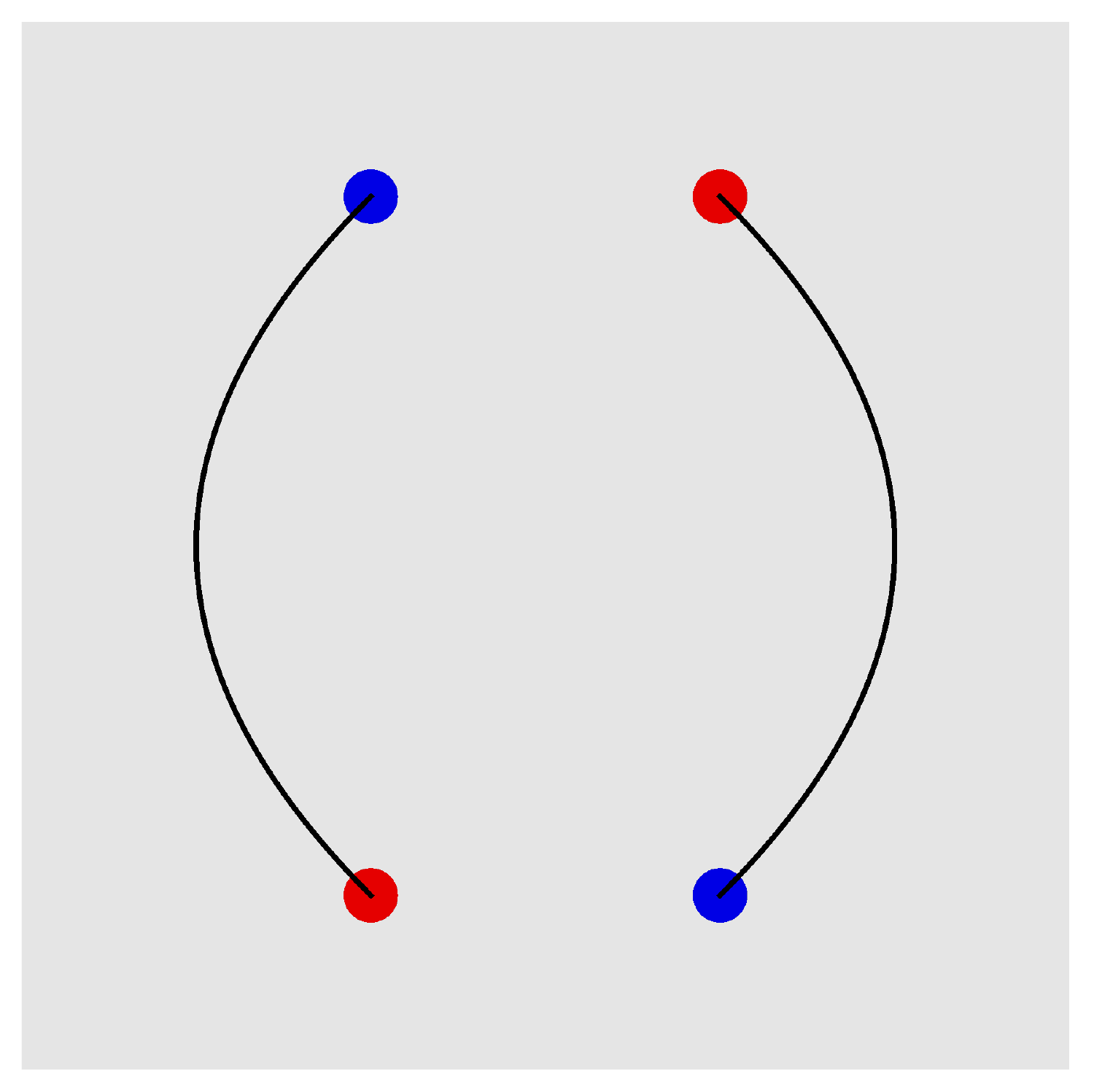}\includegraphics[scale=0.2]{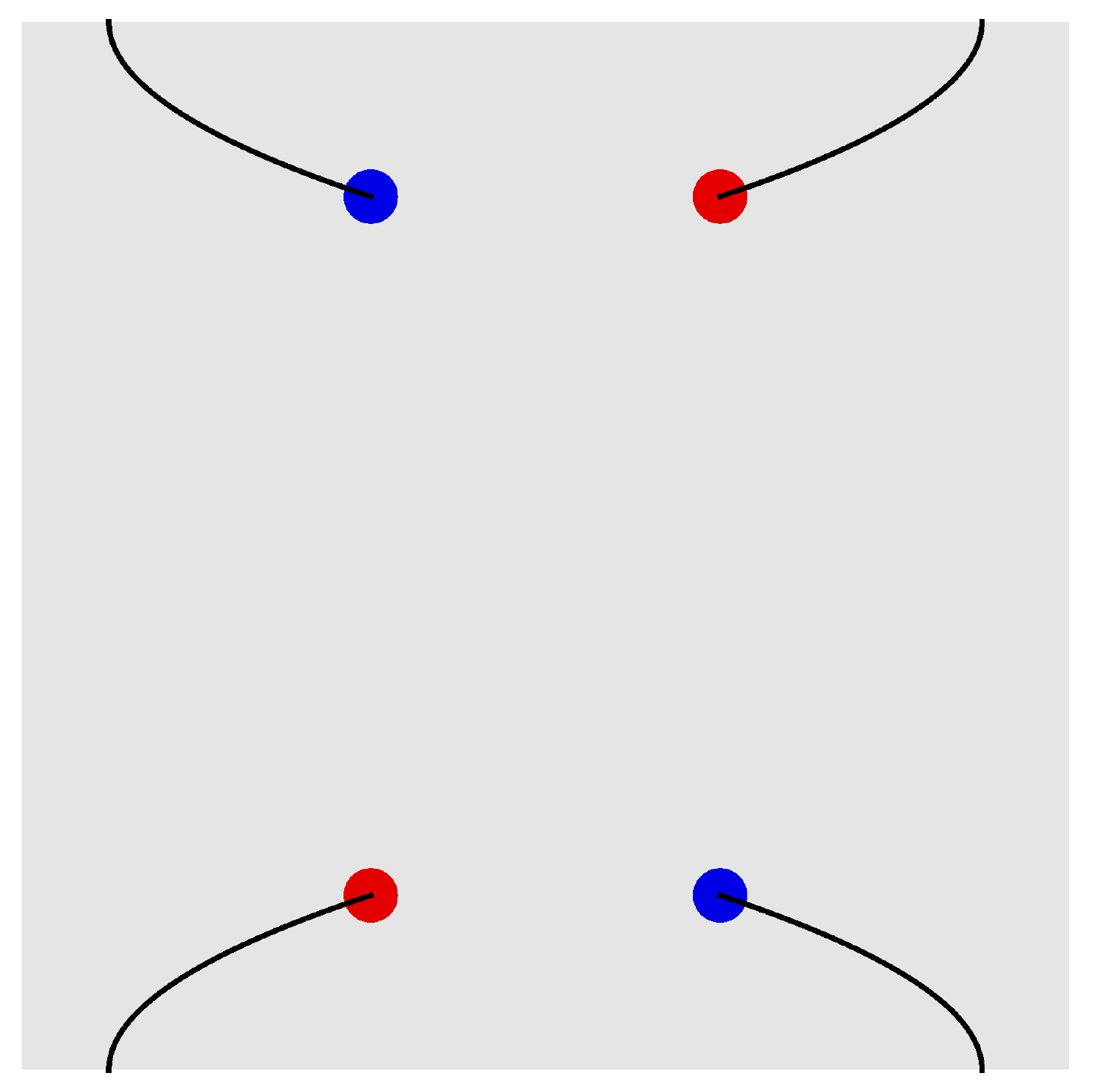}
    \caption{Possible Fermi arc connectivities for four Weyl nodes. The points indicate the surface projections of the Weyl nodes; color indicates chirality. The black lines are the Fermi arcs.}
    \label{fig:conns}
\end{figure}

\begin{figure}[h]
    \centering
    \includegraphics[scale=0.18]{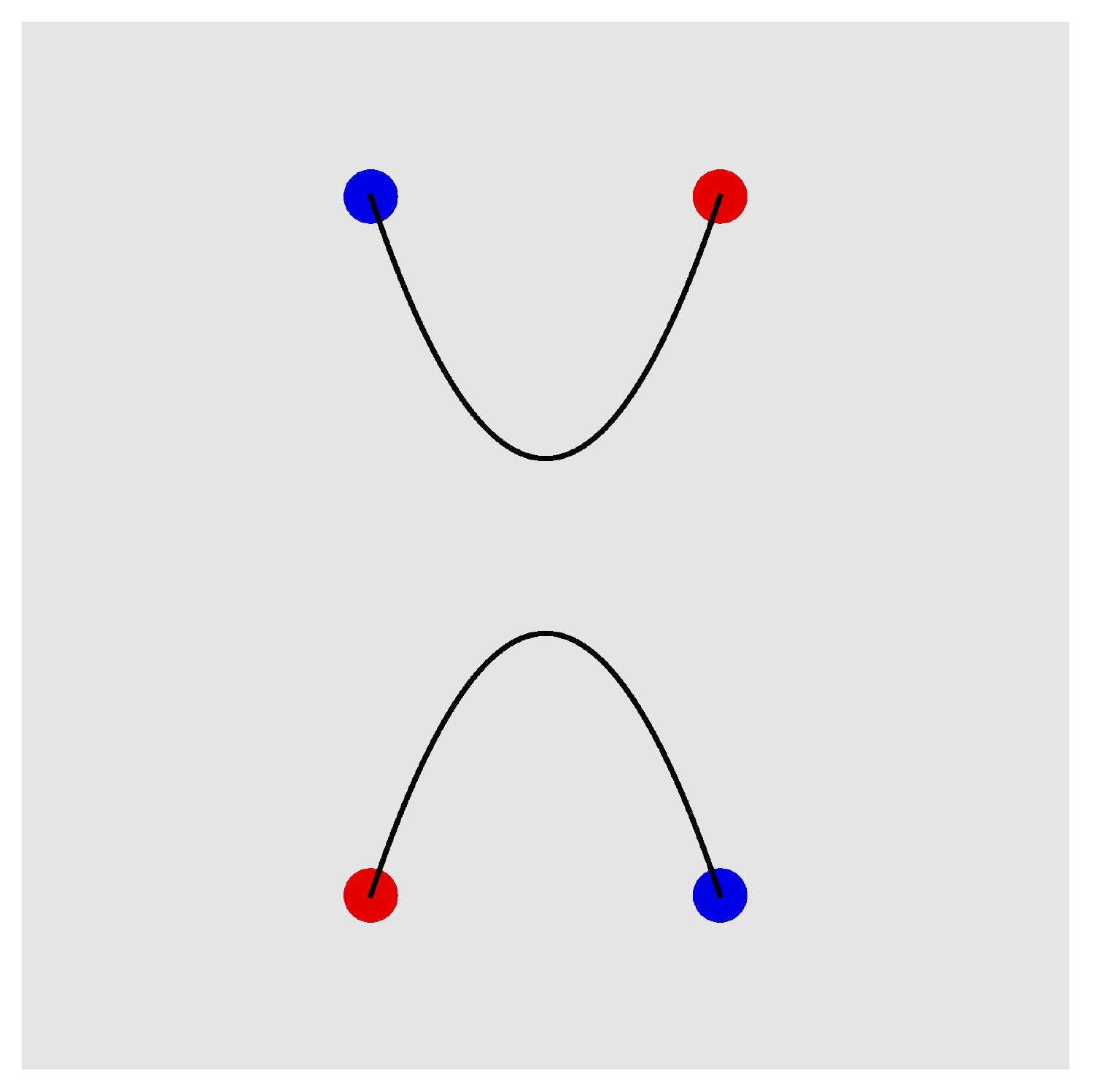}\includegraphics[scale=0.18]{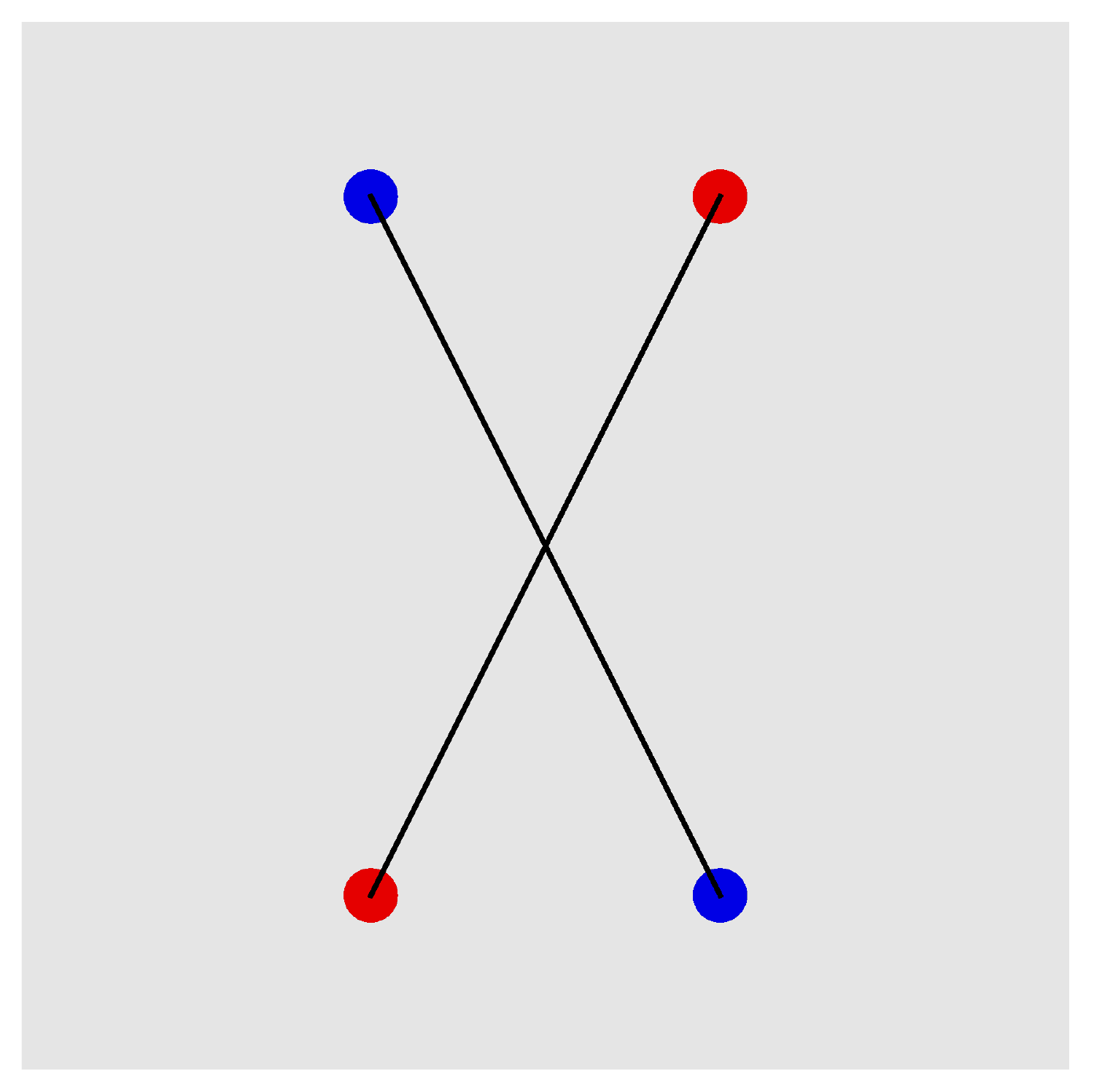}\includegraphics[scale=0.18]{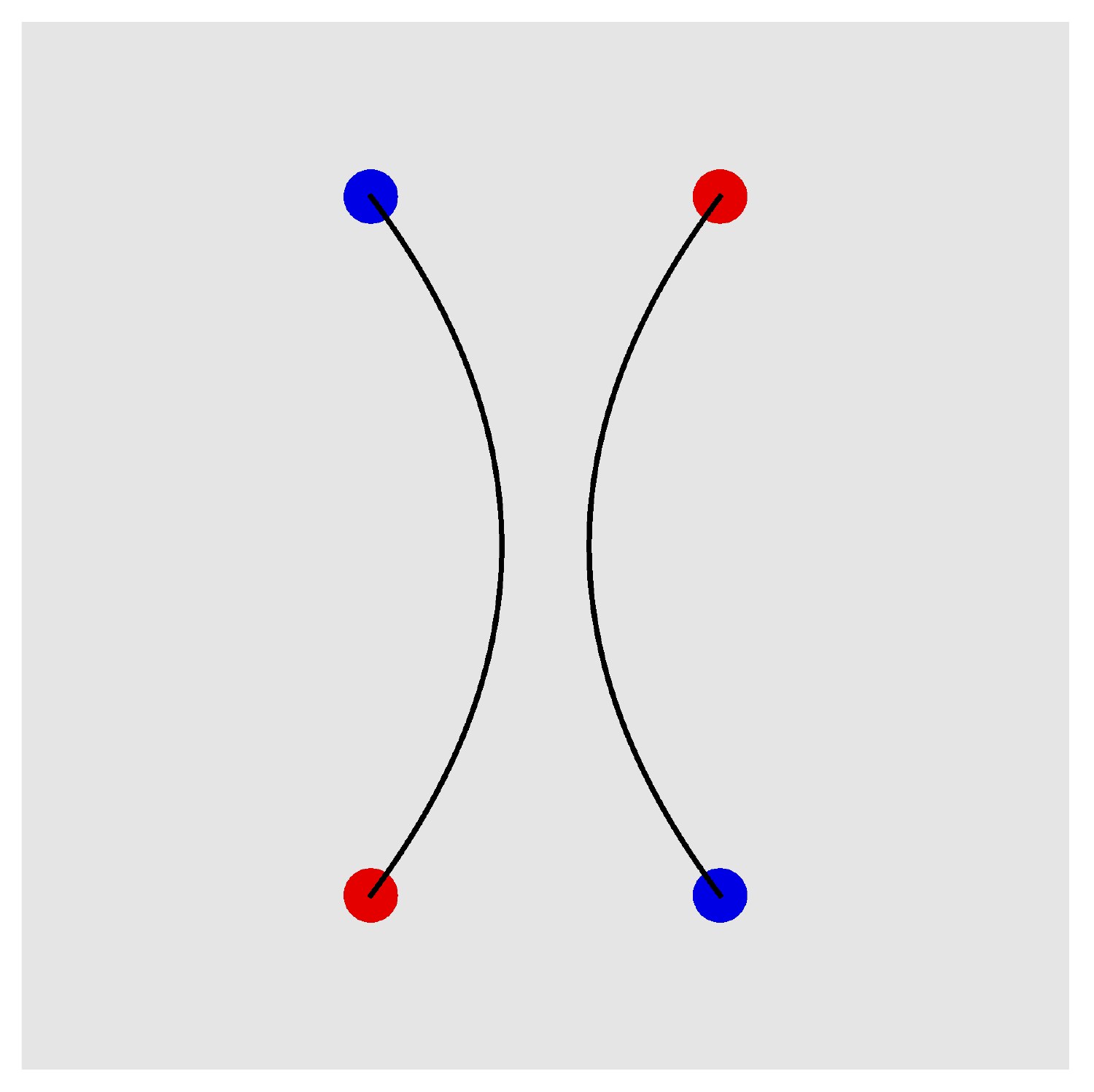}
    \caption{By smoothly transforming the surface Hamiltonian, Fermi arcs can cross and change their connectivity.}
    \label{fig:reconn}
\end{figure}

While topology may require the existence of Fermi arcs, it does not completely determine their connectivity.
In a crystal with multiple Weyl fermions, there are several possible configurations of Fermi arcs that cannot be smoothly deformed to each other, as we illustrate in Fig.~\ref{fig:conns}.
The connectivity of the Fermi arcs depends not only on the bulk Hamiltonian, but also on the details of the surface. It is possible to change the connectivity of the Fermi arcs by smoothly transforming the Hamiltonian while leaving the topology of the bulk invariant \cite{ArcCross2,WeylCrossings}, as illustrated in Fig~\ref{fig:reconn}. 
The only constraint is that the Fermi arcs satisfy the topology of the bulk Hamiltonian as well as the symmetry group of the surface Brillouin zone.
Because the surface Brillouin zone has lower symmetry than the bulk, it can be that the pattern of Fermi arcs does not satisfy bulk symmetries.
For example, in TaAs, which has a nonsymmorphic space group with fourfold screw rotations that are broken by the boundary, the surface Brillouin zone and the Fermi arcs do not have fourfold rotational symmetry \cite{TaAsTheoryArcs,TaAsArcs}.
Furthermore, because the surface Brillouin zone is not simply connected, the total momentum traversed by an arc, $\Delta\vec{k}$, is defined only up to a reciprocal lattice vector, i.e., the arcs can have nontrivial winding around the Brillouin zone.

As Fermi arcs are curves of equal energy in the two-dimensional surface dispersion relation, the velocity (i.e. gradient of energy in momentum space) of a fermion in an arc at any point is perpendicular to the arc. 
The direction of the velocity depends on the chirality of the end points.
Specifically, the velocity of a fermion in an arc is in the direction of $\hat{n}\times\  d\vec{k}$, where $\hat{n}$ is the normal to the surface from the crystal to vacuum, and $d\vec{k}$ is the change in momentum along the arc, measured from the right handed cone to the left handed cone. 
In other words, in a system with a Weyl crystal located in the half-space with $z < 0$ and vacuum where $z > 0$, the velocity of an arc emerging from a right handed cone in the direction $+\hat{x}$ is in the direction $+\hat{y}$.

\begin{figure}[h]
    \centering
    \includegraphics[scale=0.3,trim={0 3cm 0 3cm}]{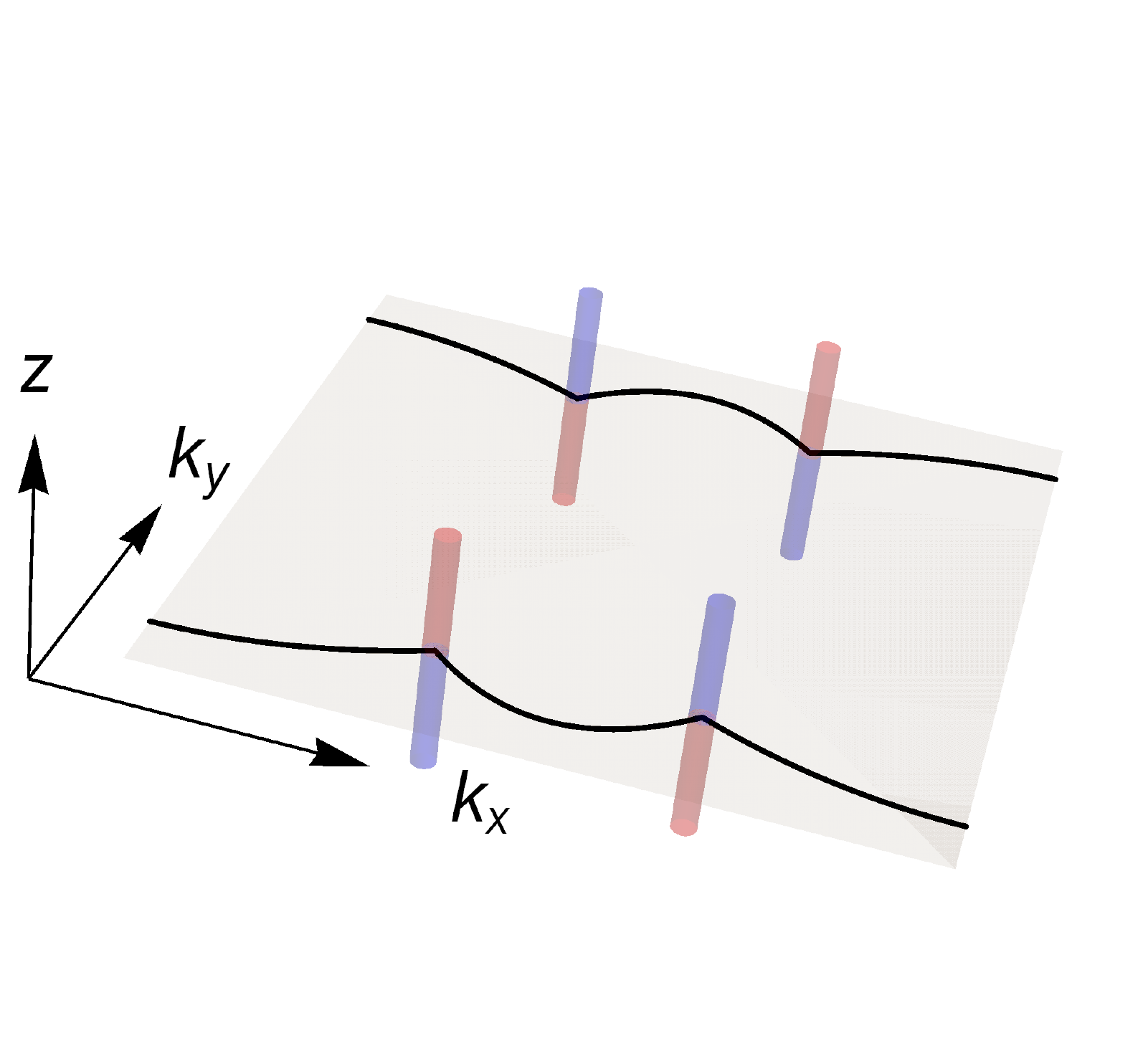}\\ \includegraphics[scale=0.3,trim={0 3cm 0 3cm}]{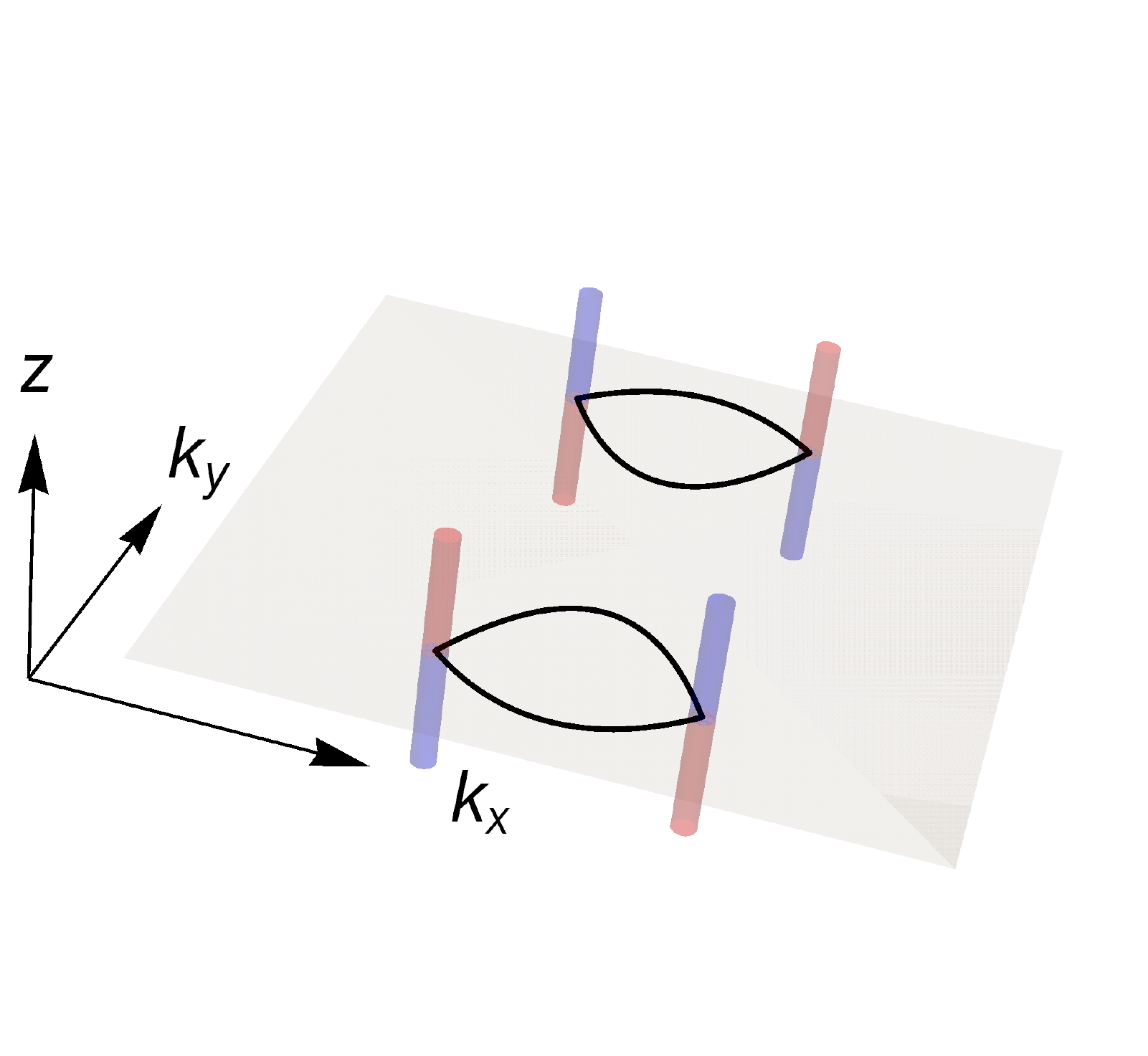}\\ \includegraphics[scale=0.3,trim={0 3cm 0 3cm}]{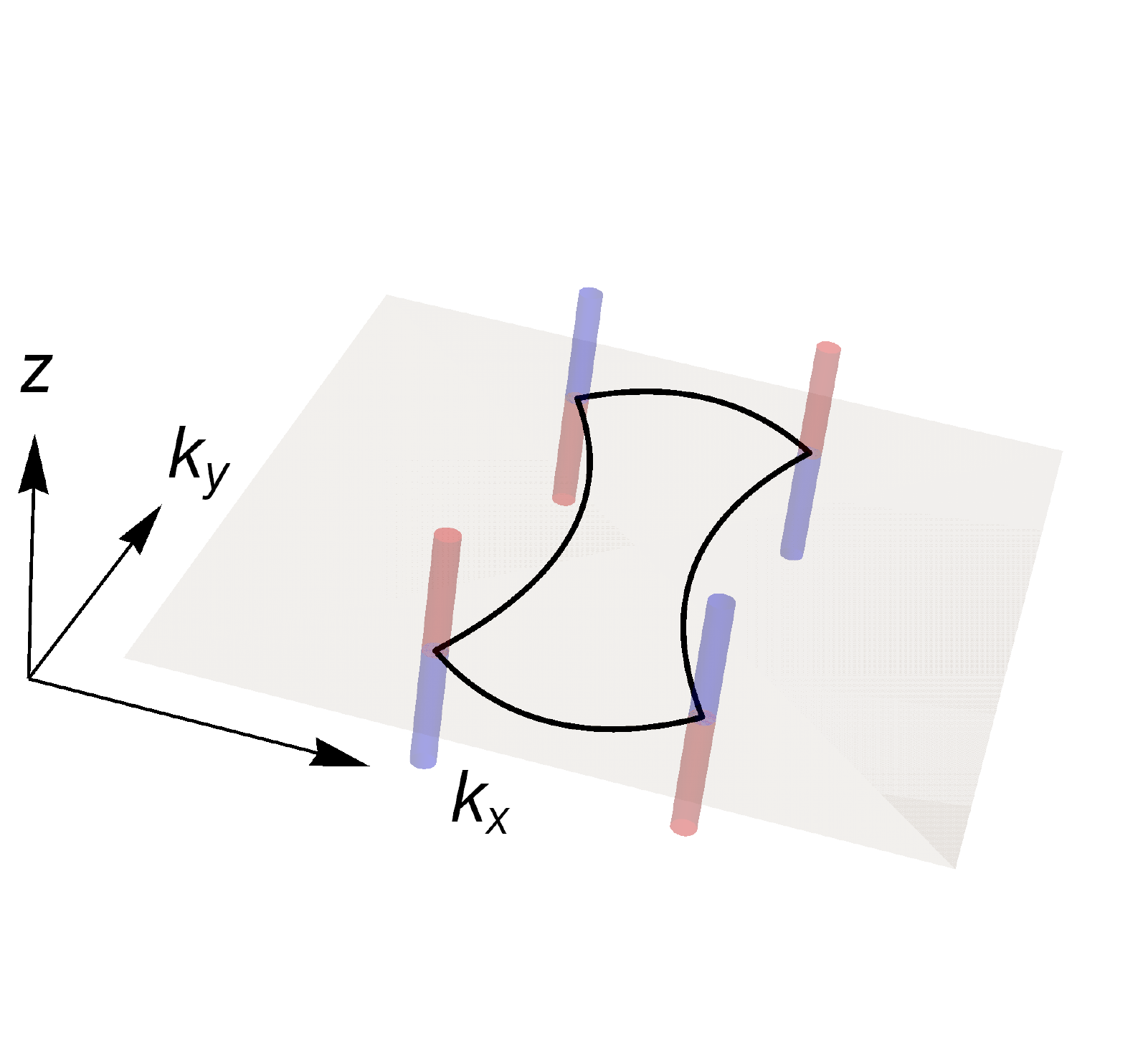}
    \caption{Possible configurations of internal Fermi arcs at a twin boundary. The sketch has momentum coordinates in plane and a position coordinate in the perpendicular direction. The Weyl cones appear as cylinders. The color indicates chirality. The grey plane is the twin boundary.}
    \label{fig:darcs}
\end{figure}

We now extend the topological argument for the existence of surface Fermi arcs to internal Fermi arcs.
Consider a twinned crystal where the projections of left handed cones of one crystal coincide with right handed cones of the other and vice versa. 
This is the case if there is mirror or glide mirror symmetry about the twin boundary. It is also the case in a time-reversal symmetric system with inversion symmetry about the twin boundary.
Now consider a loop that surrounds the projection of one Weyl point onto the boundary. If the loop is extended into a cylinder that encloses one Weyl point on each side of the twin boundary, the cylinder has opposite Chern number in the two crystals.
Therefore, the change in Chern number across the twin boundary is twice the change across a corresponding external boundary, and the projection of each cone has twice the number of Fermi arcs as the projection of the same cone at an external boundary, as sketched in Fig~\ref{fig:darcs}.

At this twin boundary, the connectivity of the arcs and the direction of velocity of the fermions within them depends not on the chirality $\chi$ of the bulk states, but on the \textit{projection} of chirality onto the boundary, $\chi \hat{n}$, where $\chi$ is the chirality of a certain bulk state, and $\hat{n}$ is the normal vector, pointing outward from the crystal that contains that bulk state. The velocity of the Fermi arc states is in the direction 
\begin{equation}\label{eq:veldir}
   \hat{v} = \chi\hat{n}\times d\vec{k},
\end{equation}
where $d\vec{k}$ is the change in momentum measured outward from a Weyl point with chirality projection $\chi\hat{n}$.

Since the Hamiltonian at the internal boundary is not the same as that of an external interface the connectivity of the internal Fermi arcs is, in general, different from two copies of external arcs. Depending on the symmetries of the boundary, it is possible that the internal arc states are completely hybridized and cannot be identified with either crystal.
Some possible connectivities are depicted in Fig~\ref{fig:darcs}.

\section{Aharanov-Bohm Phase and Quantum Oscillations}
\label{sec:QO}

We now study the dynamics of fermions at the interface in an out-of-plane magnetic field and show that transmission probability through the interface can exhibit quantum oscillations.

\begin{figure}
    \centering
    \includegraphics[scale=0.35,trim={0 2cm 0 2cm}]{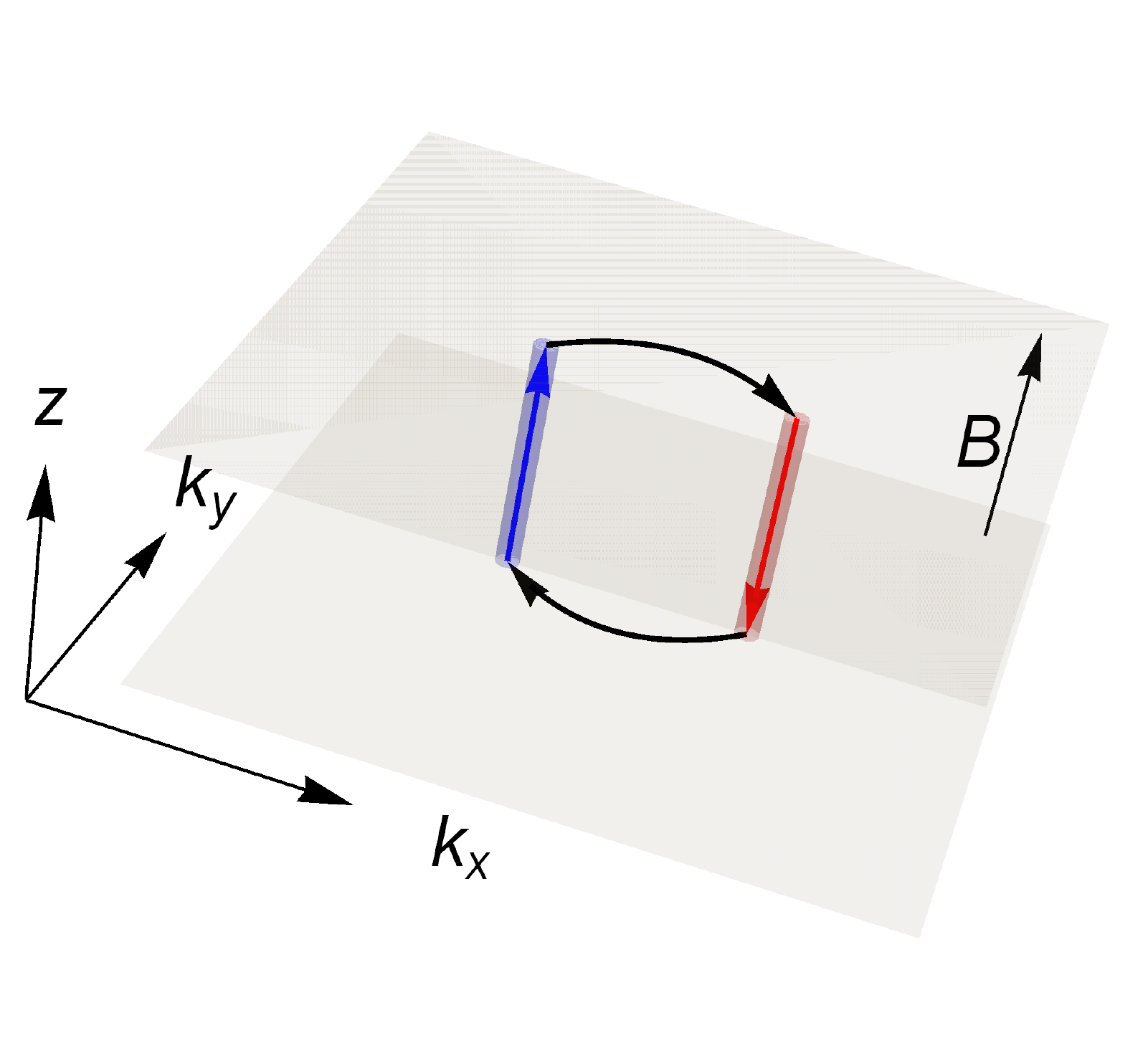}
    \caption{A closed Weyl orbit in a single crystal slab. The grey planes are the external interfaces.}
    \label{fig:simporbit}
\end{figure}

In a magnetic field, the Weyl spectrum splits into Landau levels. 
The lowest Landau level is chiral, i.e., electrons in the lowest Landau level travel only in one direction, along $\chi e \vec{B}$. 
When a fermion in the lowest Landau level reaches the boundary of a crystal, it starts travelling along the corresponding Fermi arc until it reaches a cone of opposite chirality. It then travels away from the boundary. At the opposite boundary, it similarly traverses a Fermi arc and returns to its original chirality. Such an orbit is depicted in Fig~\ref{fig:simporbit}. In this orbit, the wavefunction picks up a phase that depends on the magnetic field, the Fermi level, and the thickness of the sample, which causes quantum oscillations with a frequency different from that of oscillations from the bulk Fermi surfaces \cite{ArcOscillations1,ArcOscillations2,ArcOscillations3}. In materials with multiple Weyl cones, more complicated orbits are possible, depending on the connectivity of Fermi arcs at opposite surfaces \cite{TrefoilOrbit,WeylOrbits}. 

\begin{figure}
    \centering
    \includegraphics[scale=0.4,trim={0 3cm 0 3cm}]{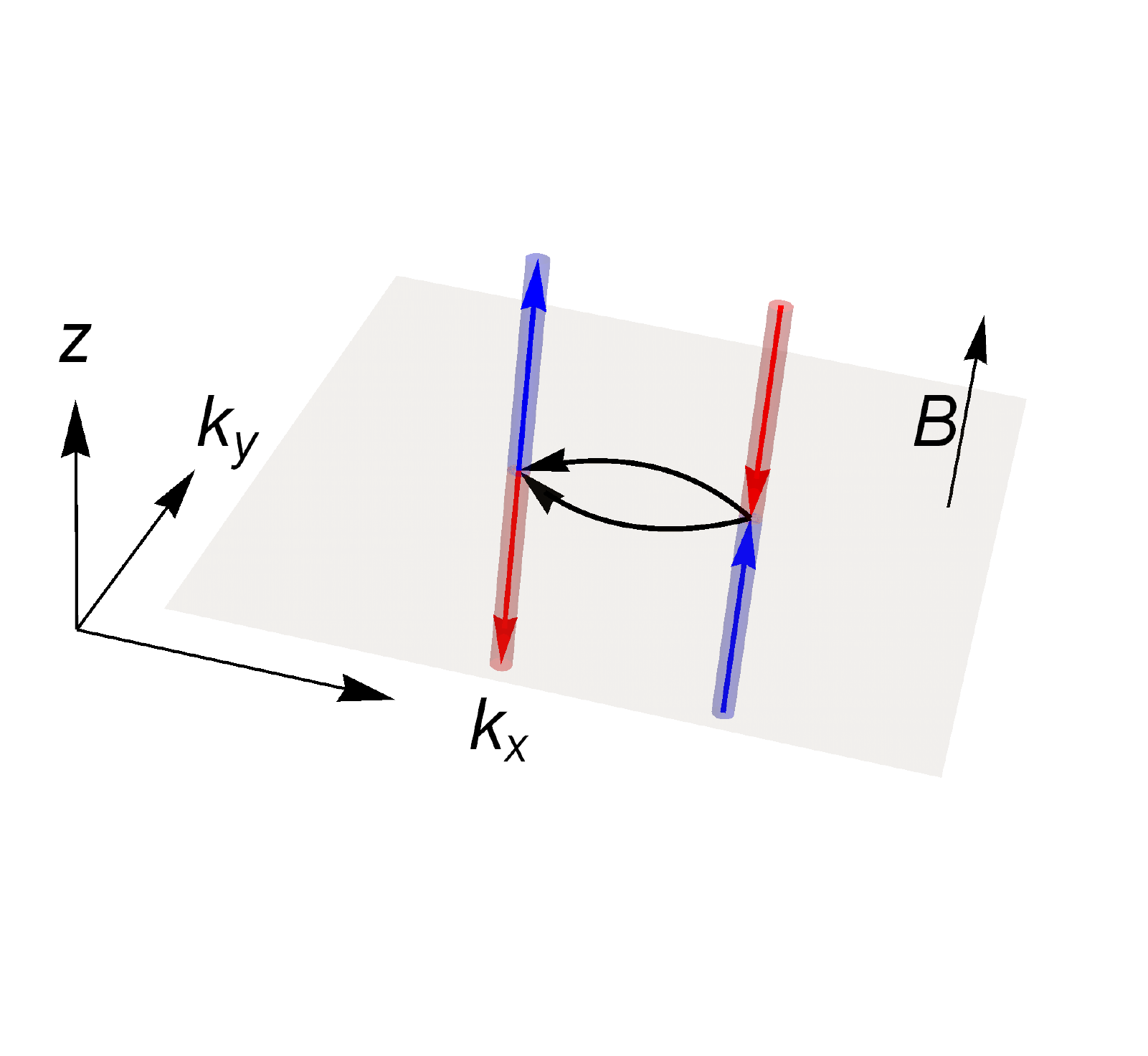}
    \caption{The path of a fermion on a twin boundary.}
    \label{fig:hybpath}
\end{figure}

At a mirror twin boundary (or inversion twin boundary with time-reversal),  there are two hybridized arcs emerging from a projection of a Weyl point, as shown in Fig.~\ref{fig:hybpath}. In the absence of magnetic field, the conductivity across the boundary will be determined by the overlap of the wavefunctions of Weyl fermions on either side \cite{GrushinInterface}. If there is a magnetic field perpendicular to the boundary, when a fermion originating from the lowest Landau level of a particular Weyl cone reaches the twin boundary, it hybridizes and travels along both Fermi arcs. After reaching the ends of the arcs, it travels away from the twin boundary either as an opposite chirality fermion in the same crystal or as a fermion of the same chirality in the opposite crystal.

Consider semiclassical dynamics of the fermions as they traverse the Fermi arcs. The fermions experience a Lorentz force; their change in momentum is perpendicular to their velocity and they move along the arcs in momentum space until they reach the other end. The change in momentum is given by:

\begin{equation}
     \frac{d}{dt} \vec{k} = \vec{v}\times  \hat{n} B_n = \frac{d}{dt} \vec{r} \times  \hat{n} B_n,
\end{equation}
where $\vec{r}$ is the position, $B_n$ is the component of the magnetic field perpendicular to the twin boundary and $\hat{n}$ is the normal vector of the boundary.
Integrating over time,
\begin{equation}
    \Delta\vec{k} = \Delta\vec{r}\times  \hat{n} B_n
\end{equation}
Since the fermions are confined to the twin boundary, this is rewritten as:

\begin{equation}
    \Delta \vec{r} = \frac{1}{B_n} \hat{n} \times \Delta\vec{k}
\end{equation}

This relation is valid everywhere along a Fermi arc, not just at its endpoints. As a fermion traverses the Fermi arc, it traces out the same shape in position space as in momentum space, but rotated by $\pi/2$ and rescaled by $1/B_n$. If two arcs share the same start and end points, and form a simply connected loop in momentum space, they have the same $\Delta\vec{k}$, and they also have the same $\Delta\vec{r}$. Such a path is illustrated in Fig~\ref{fig:hybpath}. As a fermion travels along different paths to the same endpoint, it picks up an Aharonov-Bohm phase. 
The phase difference between the paths on the two arcs is $B_n A_r = A_k/B_n$, where $A_k$ is the area traced by the arcs in momentum space, and $A_r = A_k/B_n^2$ is the area traced in position space. Depending on the Aharonov-Bohm phase, an electron originating in one crystal may be reflected back or transmitted across the boundary to the opposite crystal.
The reflection and transmission probabilities, as well as the boundary conductivity therefore have quantum oscillations with frequency $A_k$.

These quantum oscillations would occur only in a twinned crystal, not a single crystal, and would have a frequency different from oscillations caused by bulk states. These oscillations would be possible only if two internal hybridized Fermi arcs have the same start and end points and form a simply connected loop.
For example, the top configuration of Fig~\ref{fig:darcs} would not have quantum oscillations because there are pairs of arcs with the same start and end points, but the loops they form are not simply connected (and the trajectories do not have the same total $\Delta\vec{k}$ or $\Delta\vec{r}$). The bottom configuration of Fig~\ref{fig:darcs} would also not have oscillations because there are no two arcs with the same start and end points. However, the middle configuration of Fig~\ref{fig:darcs} has pairs of arcs with the same start and end points that form simply connected loops, and would exhibit quantum oscillations.

\section{Dehybridization and Interface Chiral Magnetic Effect}
\label{sec:AHE}

In magnetic Weyl materials, the interface Hall effect, which results in a transverse current if a voltage is applied to opposite surfaces of a slab, has been proposed \cite{burkov2014anomalous} and observed \cite{shekhar2018anomalous,kiyohara2016giant,nakatsuji2015large,nayak2016large,wang2018large}. 
The Hall effect occurs because the fermions in each arc have a velocity perpendicular to the change in momentum; thus, if the Fermi arcs on one side of the sample are at a higher potential than the arcs on the other side, a current is generated proportional to the length of the arcs.
Quantitatively, 
the current contributed by states in one particular Fermi arc in an energy range $\delta E$ is:
\begin{equation}\label{eq:hall}
    \vec{\kappa} = e\int \frac{1}{(2\pi)^2} v\hat{n}\times\hat{dk} \left(|dk| \frac{\delta E}{v}\right) = \frac{1}{4\pi^2} e\hat{n}\times\Delta\vec{k}\delta E
\end{equation}
In this expression, $v\hat{n}\times\hat{dk}$ is the velocity of the fermions and $|d\vec{k}| \delta E/v$ is the density of states in an infinitesimal segment of the arc with length $|dk|$ whose width in momentum space is $\delta E/v$.
Since on the opposite surface, the projection of the Weyl cone has opposite chirality, the opposite surface has opposite $\Delta\vec{k}$ for each arc. Thus, if the system is at equilibrium, the total current cancels, while if there is a potential difference between the surfaces, a current is produced with Hall conductivity $\frac{1}{4\pi^2} e\hat{n}\times\Delta\vec{k}$.

The interface Hall effect described above requires a potential difference between different Fermi arcs, which is possible if the Fermi arcs are on different surfaces and therefore spatially separated. The internal Fermi arcs are not spatially separated, so naively, we would expect the Hall current from the internal arcs alone to vanish.

However, an internal boundary current is possible if different arcs within the twin boundary have different chemical potentials. 
This requires breaking the mirror symmetry, or the combination of inversion and time-reversal symmetry, that ensures all the internal arc states have equal weight in each crystal. It also requires breaking time-reversal symmetry because chemical potential is time-reversal-even and current is time-reversal-odd. 

In the remainder of this section, we consider a small time-reversal-breaking term that could arise from applying an in-plane magnetic field or intrinsically from magnetic ordering that breaks the reflection symmetry or combination of inversion and time-reversal symmetry through the domain wall.
Since the topology of the bulk and the winding and connectivity of the Fermi arcs are robust against small perturbations, the only possible configurations of arcs in the presence of small time-reversal symmetry breaking are those that are topologically consistent with the bulk states when time-reversal symmetry is present.

\begin{figure}
    \centering
    \includegraphics[scale=0.35,trim={0 3cm 0 3cm}]{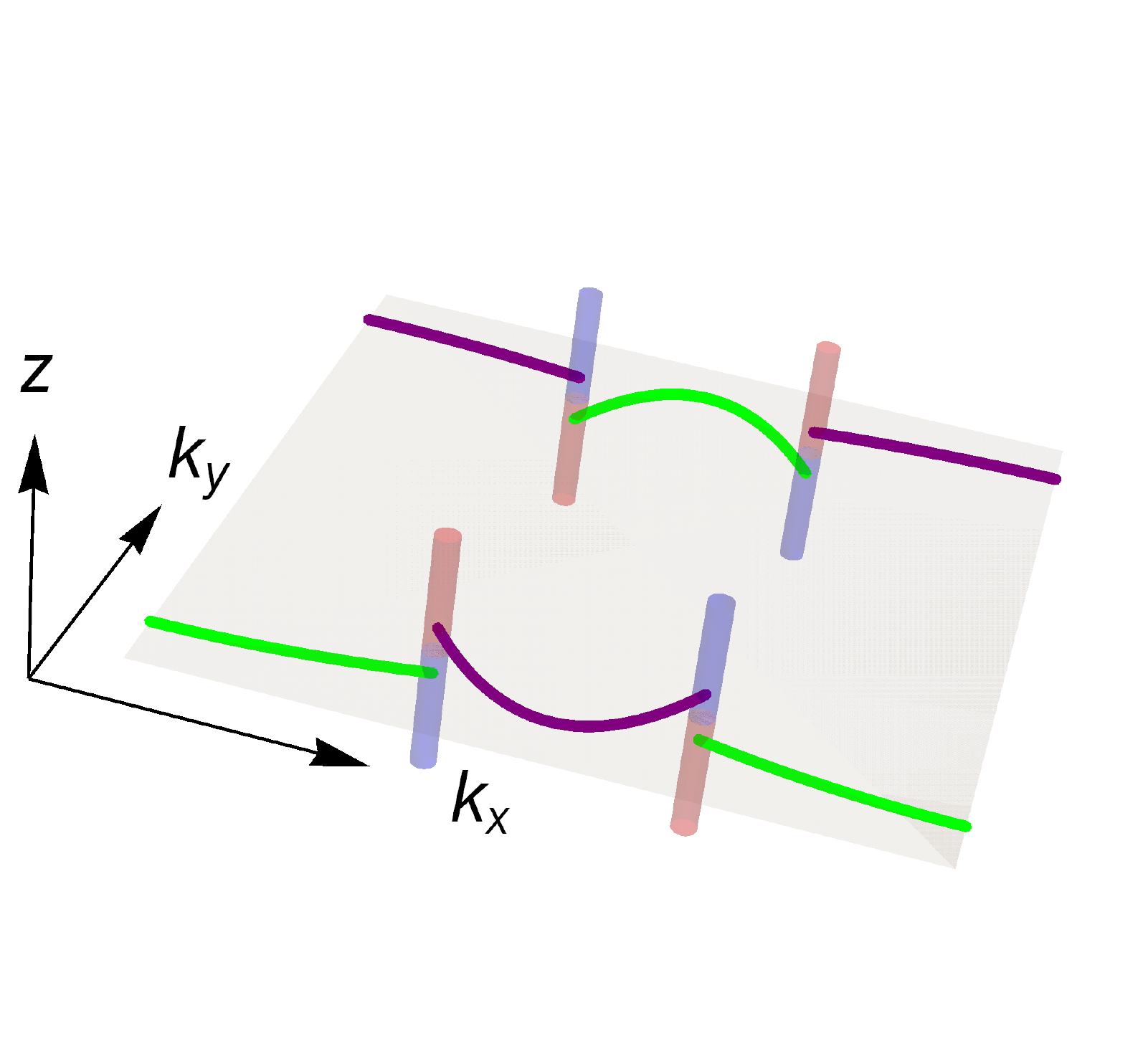}\\\includegraphics[scale=0.35,trim={0 3cm 0 3cm}]{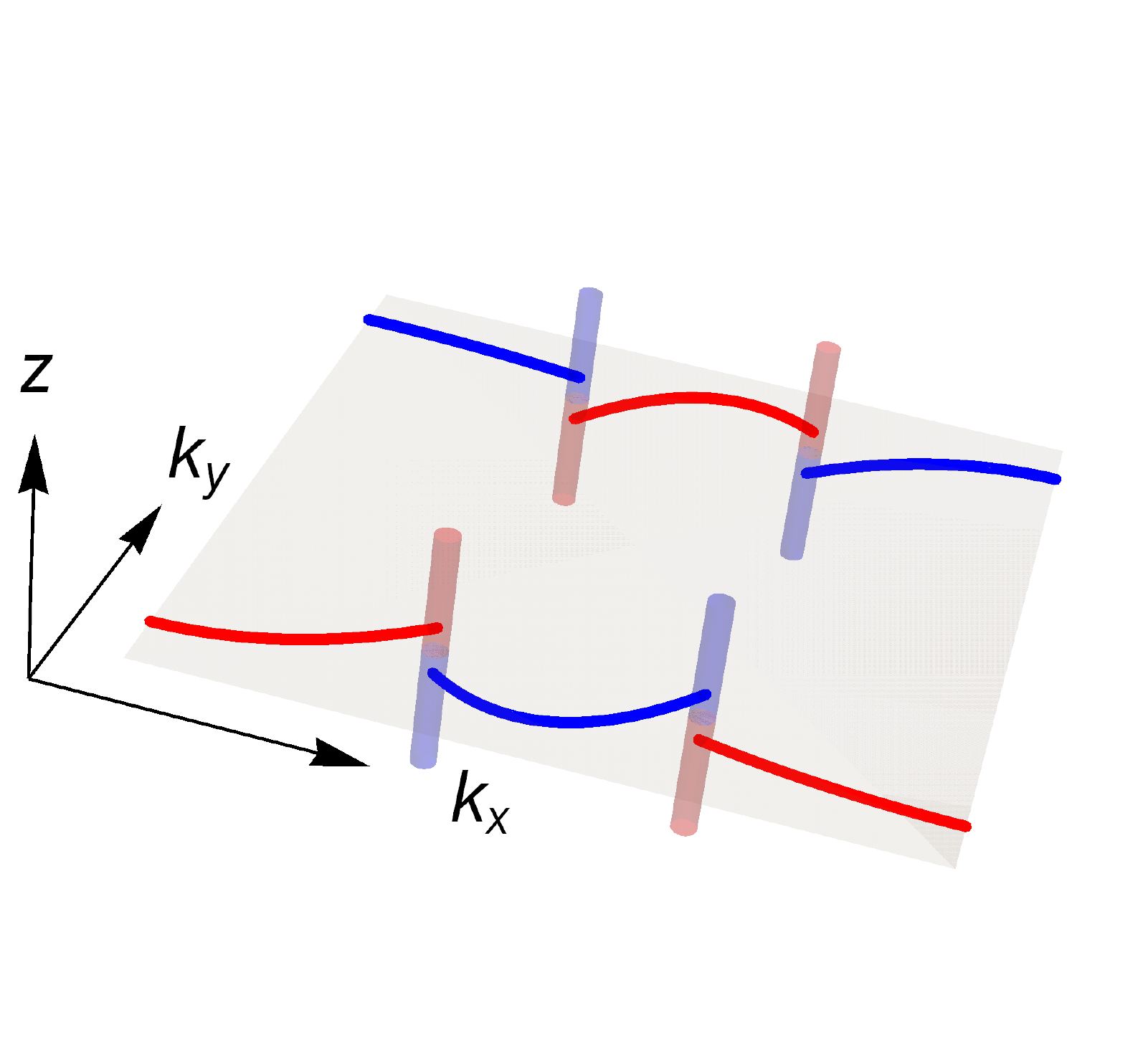}
    \caption{Dehybridized Fermi arcs. In the top picture, each arc terminates in the same crystal on both sides; thus, it can be labelled by one side of the domain wall, indicated by the purple/green color. In the bottom picture, the two end points of a single arc terminate in different crystals; thus it has definite chirality, indicated by the red/blue color.}
    \label{fig:hallarcs}
\end{figure}

Furthermore, while the Fermi arcs are localized on the twin boundary for most of their length, the ends of the Fermi arcs are delocalized and merge into the bulk spectrum.
Thus, even a small time-reversal symmetry breaking term can change the localization properties of the endpoints of each arc.
Sketches of arcs whose ends have delocalized into the bulk are shown in Fig~\ref{fig:hallarcs}.
There are two possibilities for each Fermi arc: it can have both ends delocalized into the same crystal or each end in a different crystal. We focus on the latter case, shown in the lower panel of Fig.~\ref{fig:hallarcs}. We show in section \ref{sec:toymodel} that such a configuration appears in a toy model for certain B20 materials.

When the ends of each arc are localized in different crystals, each arc connects states of the same chirality in different crystals. In this configuration, the boundary current depends on the potential difference between the left and right handed fermions, i.e. the chiral chemical potential $\mu_5$. A chiral chemical potential could be produced in both crystals by parallel electric and magnetic fields, through the chiral anomaly \cite{AnomalyAdler,AnomalyBJ,son2013chiral}.

The magnitude of the current is determined by the sum of $\Delta \vec{k}$ for all arcs whose endpoints terminate on Weyl cones with, e.g., positive chirality (the sum of $\Delta \vec{k}$ for all arcs with negative chirality must be opposite by symmetry, ensuring that there is no current in equilibrium).
If the time-reversal breaking term can be treated as a perturbation, for every cone at momentum projection $\vec{q}$, there is a cone of the same chirality projection at $\sim -\vec{q}$. Consequently, the sum of $\Delta\vec{k}$ for one set of arcs is $\sim 0$ modulo a reciprocal lattice vector; for example, in the configuration shown in the lower panel of Fig~\ref{fig:hallarcs}, the set of arcs connecting Weyl cones with positive chirality has a non-zero total $\Delta\vec{k}$.

If the fermions of opposite chirality have different chemical potentials in the bulk, the Fermi arcs corresponding to each chirality will also have different chemical potentials. This can cause a boundary current if the total $\Delta\vec{k}$ of each set is non-zero. 
Since the total $\Delta\vec{k}$ of arcs with each chirality is given by a reciprocal lattice vector, the current contributed by the twin boundary states is quantized in terms of the reciprocal lattice vectors, \textit{regardless of the distribution of Weyl cones in the bulk}. The sheet current density can be obtained by substituting $\Delta\vec{k} = \vec{G}$ into Eq.~(\ref{eq:hall}) and substituting $\delta E$ by the chiral chemical potential $\mu_5 \equiv (\mu_R - \mu_L)/2$:
\begin{equation}\label{eq:quantizedHall}
    \vec{\kappa}  = \frac{e^2}{2\pi^2} \hat{n}\times \vec{G} \mu_5.
\end{equation}
This effect is analogous to the chiral magnetic effect (CME) \cite{CME}, but it contributes a sheet current, not a bulk current, and its coefficient is quantized, while the coefficient of the bulk CME is proportional to the magnetic field.
The interface chiral magnetic effect in Eq.~(\ref{eq:quantizedHall}) cannot occur in a single crystal because the Fermi arcs cannot have the property where both ends are localized to a Weyl cone of the same chirality.

\begin{figure}
    \centering
     \includegraphics[scale=0.35,trim={0 3cm 0 3cm}]{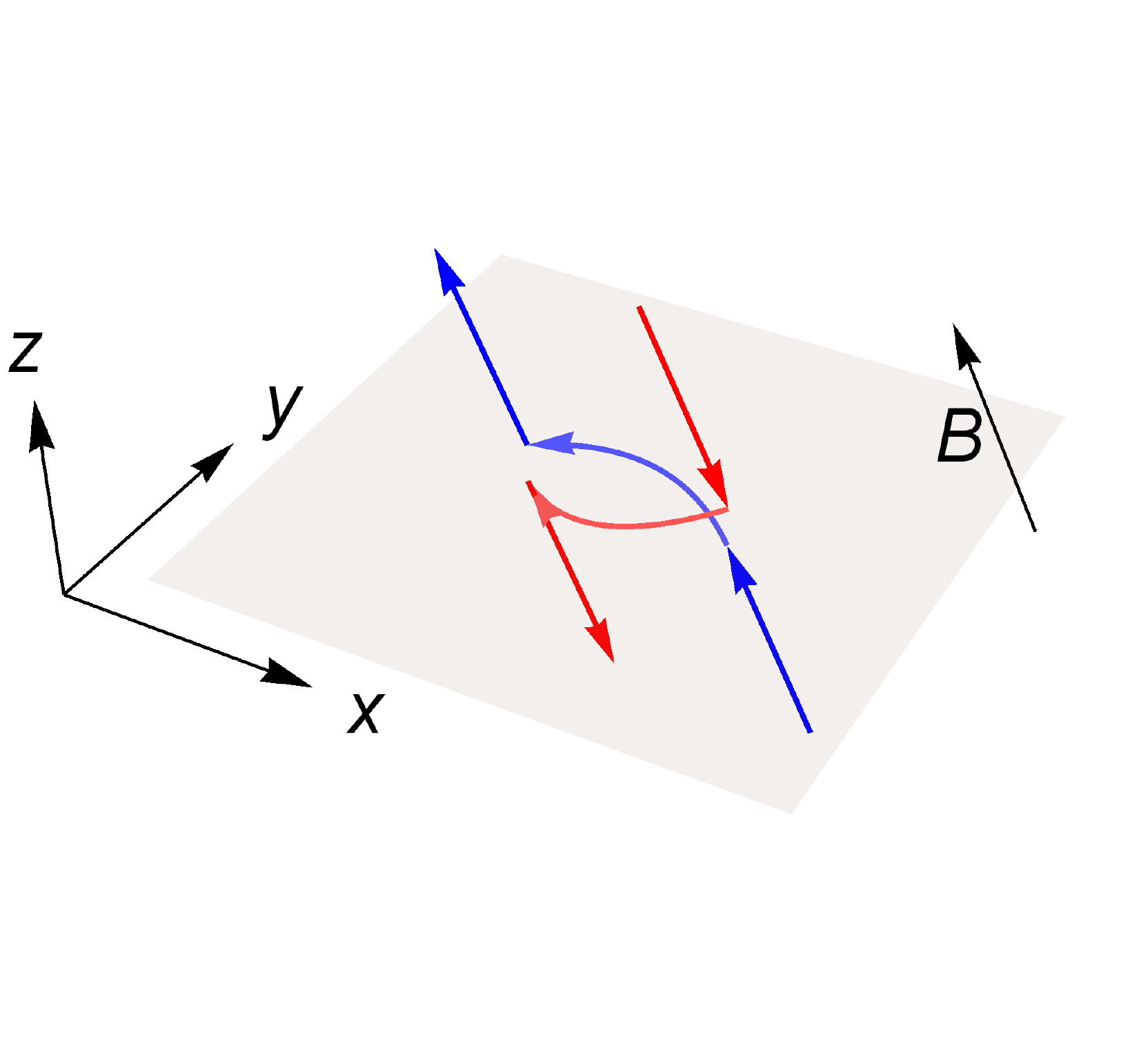}\\\includegraphics[scale=0.35,trim={0 3cm 0 3cm}]{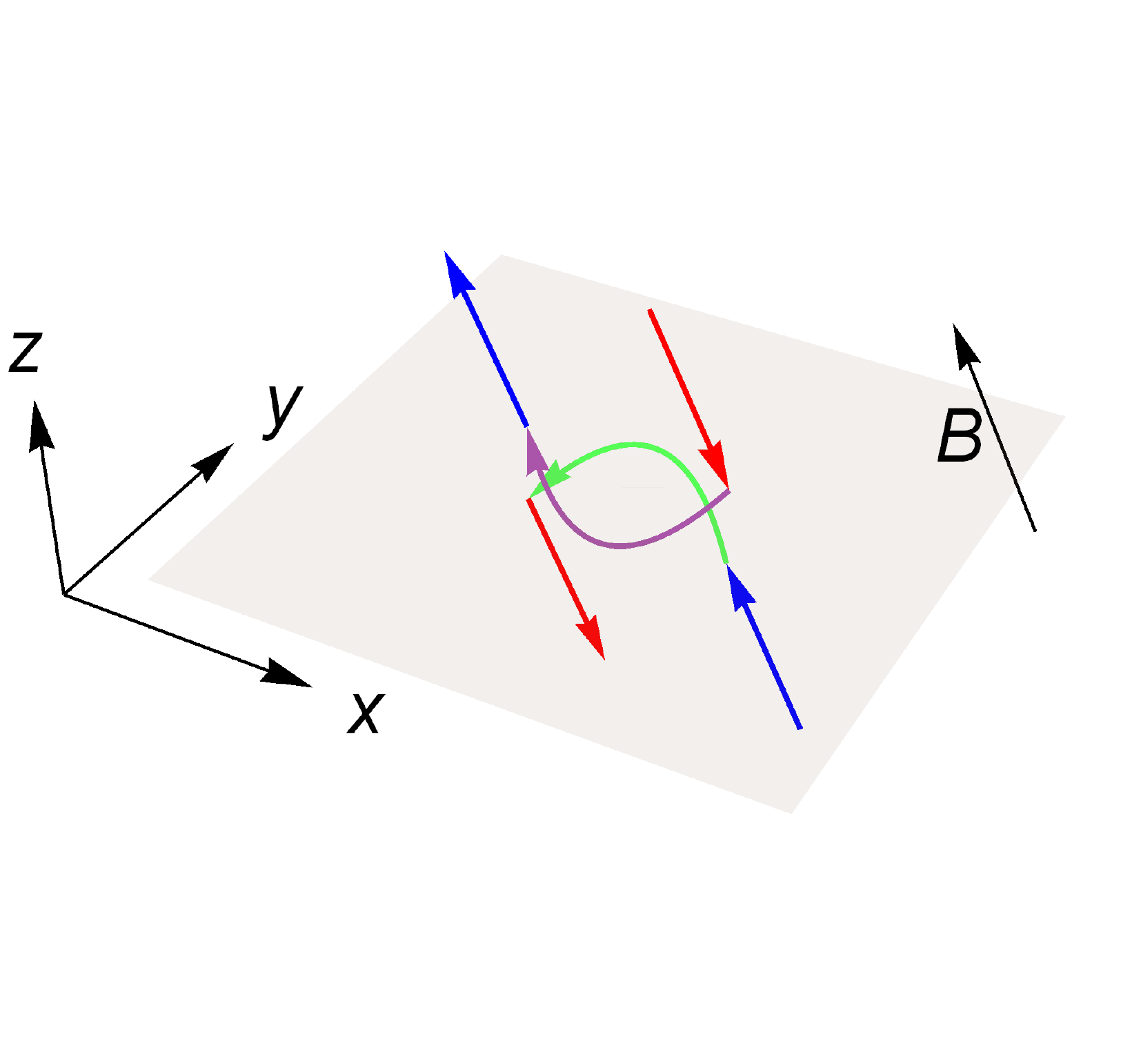}
    \caption{Semiclassical trajectories in an oblique magnetic field. Red and blue indicate chirality; green and purple indicate the two crystals.}
    \label{fig:ObliqueField}
\end{figure}

We now consider an oblique magnetic field instead of an in-plane field. The in-plane component will cause the ends of the Fermi arcs to dehybridize and the out-of-plane component will cause fermions to move along the arcs. In the bulk, the fermions would have oblique trajectories parallel to the magnetic field. As they approach the twin boundary, the fermions in the lowest Landau level of a particular Weyl cone transfer to the arc connected to that cone. They traverse the arc and transfer to the cone that is connected to the opposite end of the arc. Such trajectories are sketched in Fig~\ref{fig:ObliqueField}. 
If the dehybridized arcs connect cones of same chirality in opposite crystals, the fermions will be transmitted and the conductivity across the interface will be enhanced; if they connect cones of opposite chirality in the same crystal, the fermions will be reflected and the conductivity will be suppressed. Therefore, in an oblique field, the transmission or reflection will depend on the dehybridization of the arcs (determined by the in-plane component of the field), but not on the relative phase between two paths (determined by the out-of-plane component).
Thus, the conductivity across the boundary will depend strongly on the in-plane component of the magnetic field, and will not have quantum oscillations in the out-of-plane component.

\section{Material Candidates: Twinned B20 Crystals}\label{sec:B20}

We now apply our results to Weyl materials in the B20 family of chiral crystals, and demonstrate that they can host internal Fermi arcs at twin boundaries. Therefore, they are possible candidates to observe the effects predicted above. We are interested in this family of materials because they exhibit long Fermi arcs that span the entire Brillouin zone \cite{tang2017multiple,chang2017unconventional,schroter2019chiral,schroter2020observation} and because merohedrally twinned B20 crystals have already been grown as nanostructures and studied for spintronic applications \cite{szczech2010epitaxially, NW1,NW2,NW3,NW4}.

\begin{figure}
    \centering
    \includegraphics[scale=0.25]{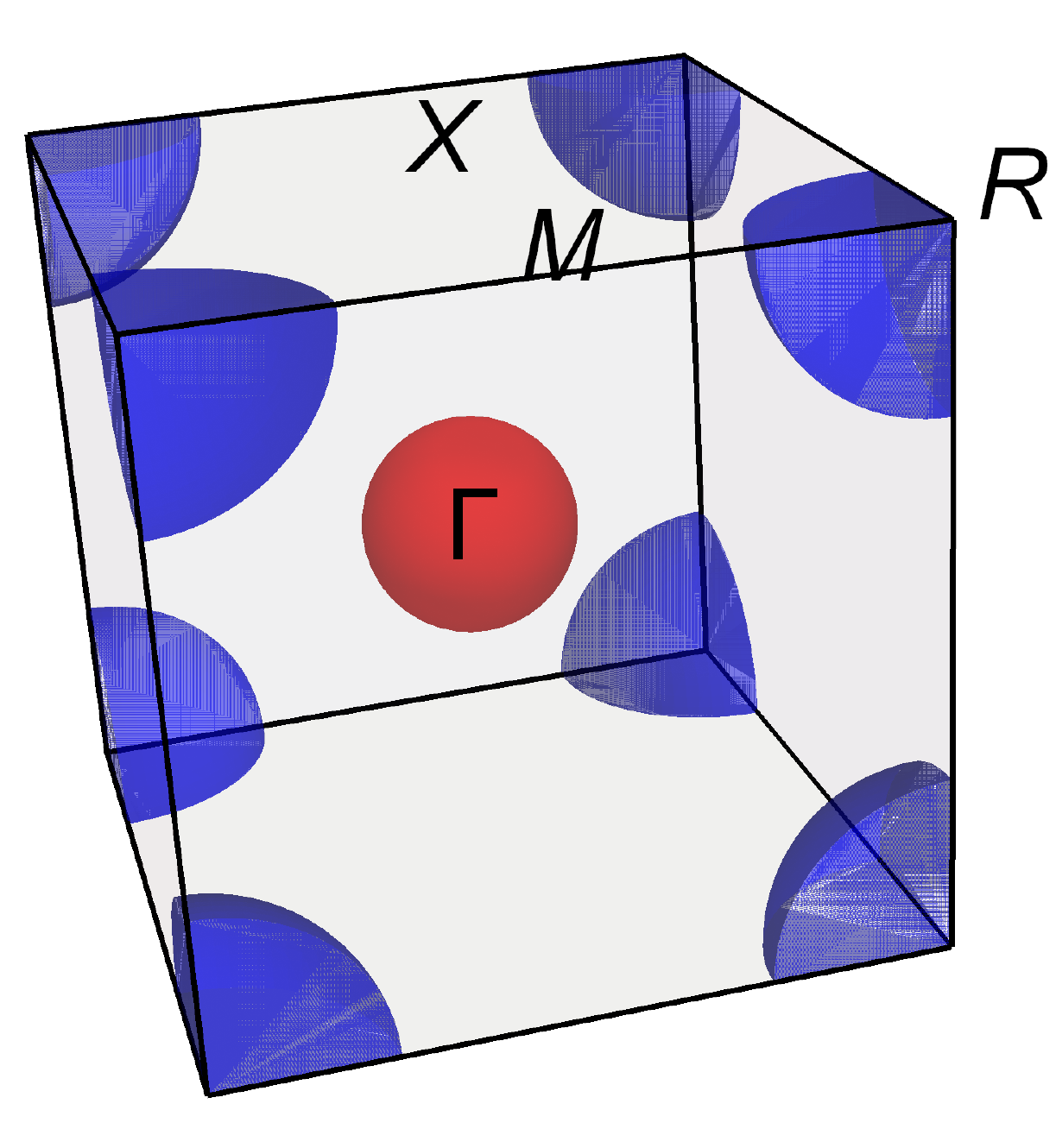}\hspace{0.05\textwidth}\includegraphics[scale=0.3]{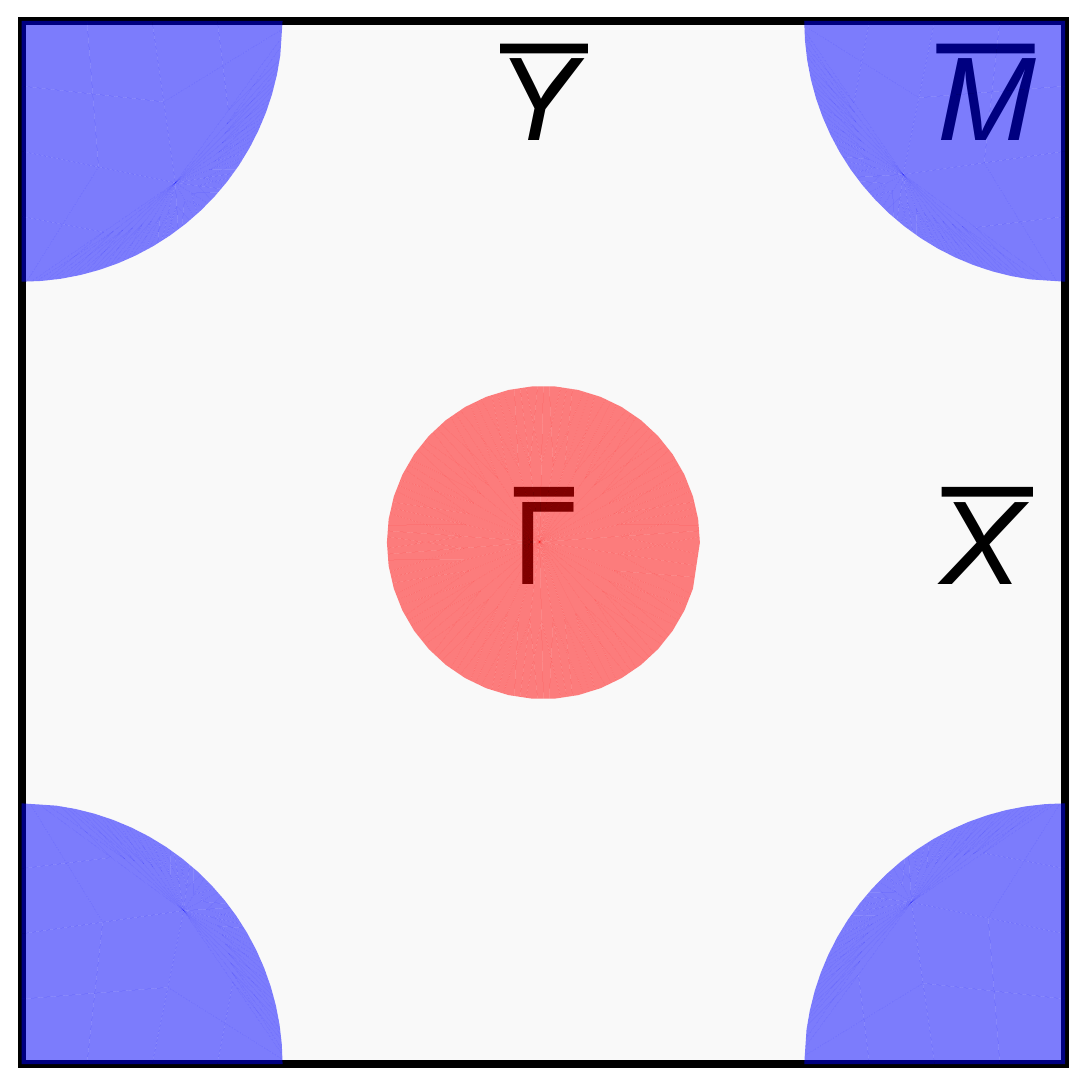}
    \caption{The bulk Fermi surfaces of a generic B20 crystal with chiral fermions and their projection onto a (001) plane (Fermi arcs not shown). The color denotes chirality. Note that $\bar{M}$ is the projection of $R$. Also, $\bar{X}$ and $\bar{Y}$ are not equivalent.}
    \label{fig:BZ}
\end{figure}

B20 materials with chiral fermions and Fermi arcs include AlPt~\cite{schroter2019chiral}, PdGa~\cite{sessi2020handedness,schroter2020observation}, CoSi~\cite{takane2019observation,rao2019observation,sanchez2019topological,xu2020optical,ni2021giant} and RhSi~\cite{sanchez2019topological,ni2020linear}. They have a simple cubic unit cell and host chiral multifold fermions at $\Gamma = (0,0,0)$ and $R=(\pi,\pi,\pi)$ that each have Chern number $\pm 4$. Ignoring spin-orbit coupling, there is a spin-1 fermion at $\Gamma$ and a double spin-1/2 fermion at $R$. Spin-orbit coupling splits the node at $\Gamma$ into a spin-3/2 and a spin-1/2 fermion, and the node at $R$ into a double spin-1 fermion and a 
trivial two-fold degeneracy
\cite{CanoMultifold,FlickerMultifold}. 
The bulk Fermi surfaces and their projection onto the $(001)$ plane are sketched in Fig~\ref{fig:BZ}.

Consequently, there are four spinful Fermi arcs emanating from the surface projection of each multifold fermion.
Since $\Gamma$ and $R$ are maximally separated in momentum space, the Fermi arcs span the entire surface Brillouin zone. 
Further, because $\Gamma$ and $R$ are not related by crystal symmetries, these materials can exhibit phenomenology not possible in symmetric Weyl materials such as TaAs, including quantized circular photocurrents \cite{deJuanQCPE,FlickerMultifold,RhSiCurrent,ni2021giant}, a chiral magnetic effect induced by strain \cite{VozmedianoStrain} or AC voltage \cite{YutaCME}, and the helical magnetic effect \cite{YutaHME}, which can produce a magnetic photocurrent in response to linearly polarized light \cite{SahalHME}.

The B20 materials are binary compounds with the nonsymmorphic space group $P2_13$ (198) and point group $T$.  
The simple cubic space group contains threefold rotations about axes $\langle 111\rangle$, and twofold screw rotations (with half translations) about axes $\langle 100\rangle$ \cite{BilBao1,Bilbao2,Bilbao3,BilbaoServer}.

Each unit cell contains four atoms of each species, residing at the fourfold Wyckoff positions shown in Table~\ref{tab:B20pos}.
\begin{table}[b]
\centering
\begin{tabular}{c|c} 
 \hline
 Atom A & Atom B\\
 \hline\hline
$(x,x,x)$ & $(-y,-y,-y)$\\ $(1/2-x,-x,1/2+x)$ & $(1/2+y,y,1/2-y)$\\ $(-x,1/2+x,1/2-x)$ & $(y,1/2-y,1/2+y)$\\ $(1/2+x,1/2-x,-x)$ & $(1/2-y,1/2+y,y)$\\
 \hline
\end{tabular}
\caption{Atomic positions for a B20 crystal.}
\label{tab:B20pos}
\end{table}
In an ``ideal" B20 structure, where $x$ and $y$ are exactly $(\sqrt{5}-1)/8 \approx 0.15451$, each atom has seven nearest neighbors of the opposite species that lie at the vertices of a regular dodecahedron, and six next-nearest neighbors of the same species \cite{B20a,B20b,B20c}. In real crystals, these parameters have slightly different values, resulting in the seven ``nearest" neighbors having slightly different distances, but all six next-to-nearest neighbors remaining equidistance. 

\begin{figure}
    \centering
    \includegraphics[scale=0.2]{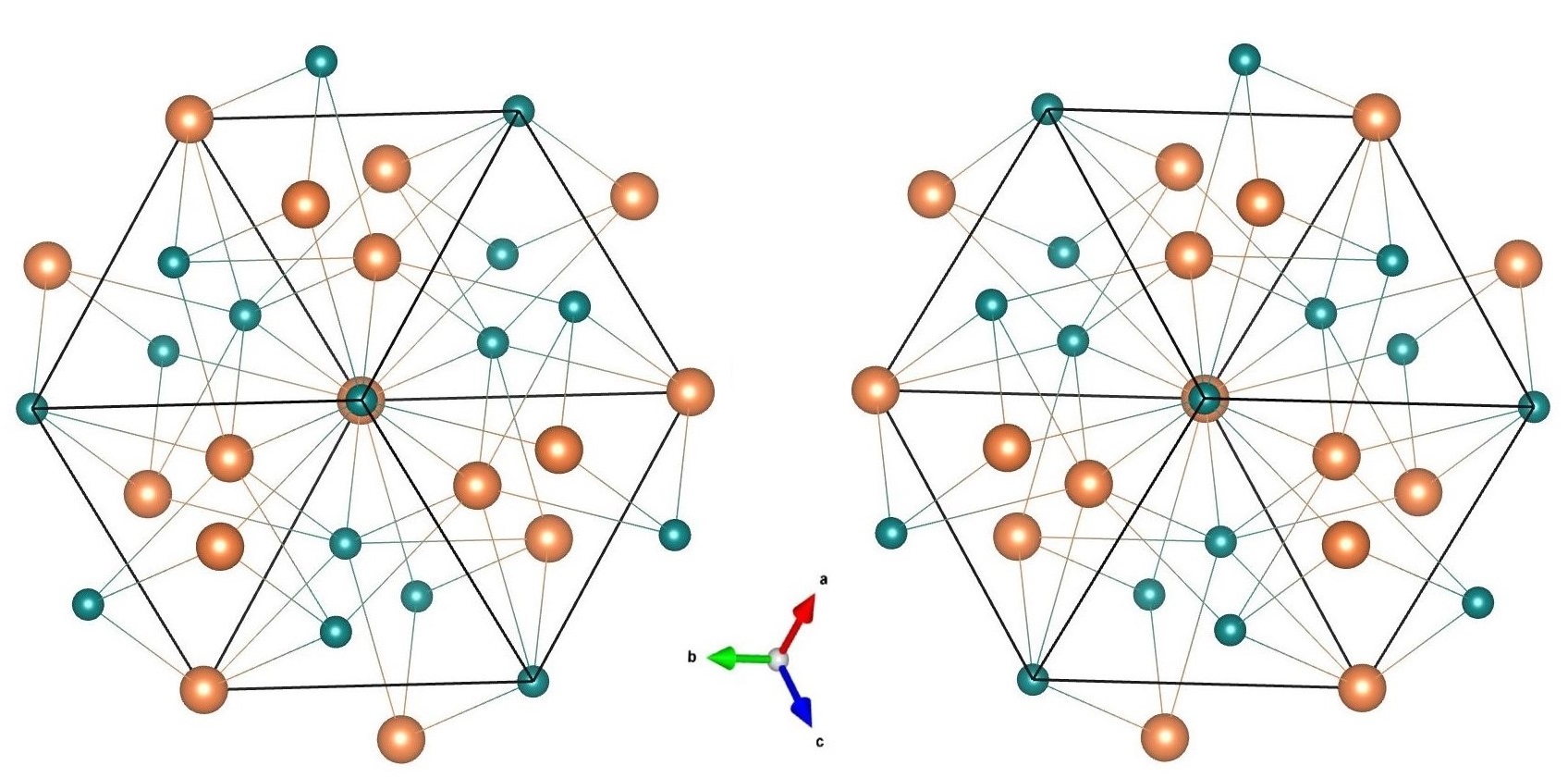}
    \caption{Crystal structures of left and right handed B20 crystals looking down $\langle 111\rangle$. Figure created using VESTA \cite{VESTA}.}
    \label{fig:B20}
\end{figure}

The B20 materials exist in two enantiomers, left- and right-handed, as illustrated in Fig.~\ref{fig:B20}.
The multifold fermions in the two enantiomers have opposite Berry curvature, i.e., at the same high-symmetry point, a charge-four source of Berry curvature in one enantiomer corresponds to a charge of negative four in the other.
Further, their Fermi arcs are mirror images of each other, which has been observed via angle resolved photo emission \cite{schroter2020observation} and scanning tunneling microscopy \cite{sessi2020handedness}. For some B20 materials, such as FeSi and MnSi (which do not have chiral fermions), merohedrally twinned structures exhibiting crystals of opposite chirality separated by a twin boundary have been synthesized and structurally characterized. \cite{szczech2010epitaxially,NW1,NW2,NW3,NW4}.

A twin boundary between the two enantiomers will host eight (counting spin) gapless Fermi arcs localized on the twin boundary, as pointed out in Ref.~\cite{schroter2020observation}.
Each surface contributes four gapless Fermi arcs, but the Fermi arcs cannot annihilate due to the multifold fermions at the same high-symmetry point having opposite Berry curvature in the two enantiomers, and therefore same chirality projection, as we argued earlier for Weyl fermions.

\begin{figure}
    \centering
 \includegraphics[scale=0.2]{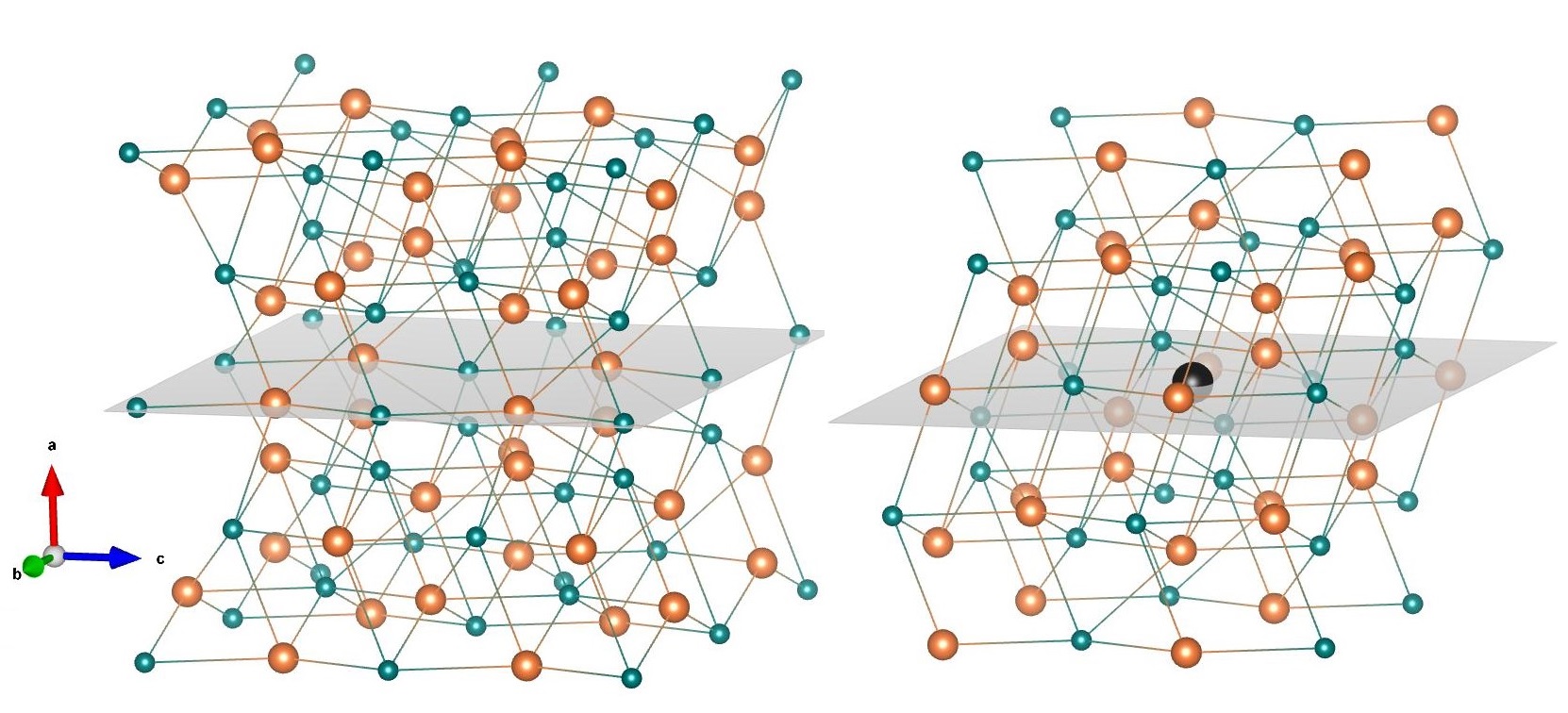} 
    \caption{Mirror (left) and inversion (right) twinned configurations of a B20 material. The grey plane is the twin boundary. The black dot in the lower figure is the inversion center. Figure created using VESTA \cite{VESTA}.}
    \label{fig:B20twins}
\end{figure}

We focus on merohedral twinning (i.e. twinning where both crystals have the same lattice vectors) with a $(001)$ twinning plane. There are two possible configurations: 
one in which the two crystals share a plane and are related by mirror symmetry about that plane, and a second in which the crystals share two planes, and are related by inversion symmetry with a center between the planes.
Both configurations are depicted in Fig.~\ref{fig:B20twins} for an ``ideal'' B20 crystal where nearest-neighbor and next-nearest neighbor distances are preserved.
The inversion-twinning configuration also preserves the coordination number for each atom, while the mirror-twinning configuration does not.



\section{Simplified Model for twinned B20 crystals}\label{sec:toymodel}

\begin{figure}
    \centering
    \includegraphics[scale=0.25]{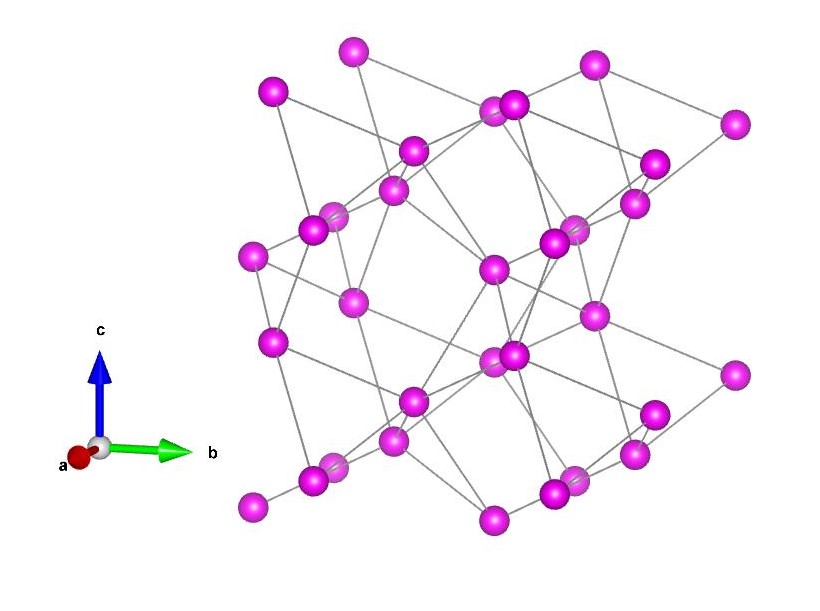}
    \caption{Crystal structure of one sublattice of the simplified model. Figure created using VESTA \cite{VESTA}.}
    \label{fig:ToyModel}
\end{figure}

To model the B20 crystals, we consider a more symmetric crystal whose atoms reside at Wyckoff positions with $x=y=1/8$ in Table~\ref{tab:B20pos}. In this structure, the atoms are displaced by $\sim 0.05 a$ compared to the ``ideal" B20 structure described in the previous section. 
The sublattice formed by atoms of one element (shown in Fig~\ref{fig:ToyModel}) has the symmetry of the space group $P4_3 32$ (SG 212), while the other sublattice has the symmetry of space group $P4_1 32$ (SG 213).
These space groups are enantiomorphs of each other: under reflection or inversion, the lattice in space group $P4_3 32$ transforms into the lattice in space group $P4_1 32$ and vice versa. 
Both space groups have the point group $O$, corresponding to a simple cubic unit cell, and contain the space group of the B20 crystal, $P2_1 3$ (SG 198), as a subgroup.

We now consider a tight-binding model that describes hopping between atoms on only one sublattice, which will describe the low-energy physics when the bands near the Fermi level originate mostly from one atom.
The model ignores spin-orbit coupling and contains only nearest neighbor hopping, given by the parameter $t$.  The distance between nearest neighbors is $\sqrt{3/8}a \sim 0.612a$, where $a$ is the lattice constant. 
Explicitly, the bulk Hamiltonian for the sublattice in space group $P4_3 32$ is given by: 
\begin{widetext}
\begin{equation}
    H = 2t\begin{pmatrix}
    0 & \cos(k_y a/2) e^{i(k_z - k_x)a/4} & \cos(k_z a/2) e^{i(k_x - k_y)a/4} & \cos(k_x a/2) e^{i(k_y - k_z)a/4}\\
    \cos(k_y a/2) e^{i(-k_z + k_x)a/4} & 0 & \cos(k_x a/2) e^{i(k_y + k_z)a/4} & \cos(k_z a/2) e^{i(-k_x - k_y)a/4}\\
    \cos(k_z a/2) e^{i(-k_x + k_y)a/4} & \cos(k_x a/2) e^{i(-k_y - k_z)a/4} & 0 & \cos(k_y a/2) e^{i(k_z + k_x)a/4}\\ 
    \cos(k_x a/2) e^{i(-k_y + k_z)a/4} & \cos(k_z a/2) e^{i(k_x + k_y)a/4} & \cos(k_y a/2) e^{i(-k_z - k_x)a/4} & 0
    \end{pmatrix}
\label{eq:Ham}
\end{equation}
\end{widetext}

\begin{figure}
     \centering
     \includegraphics[scale=0.27]{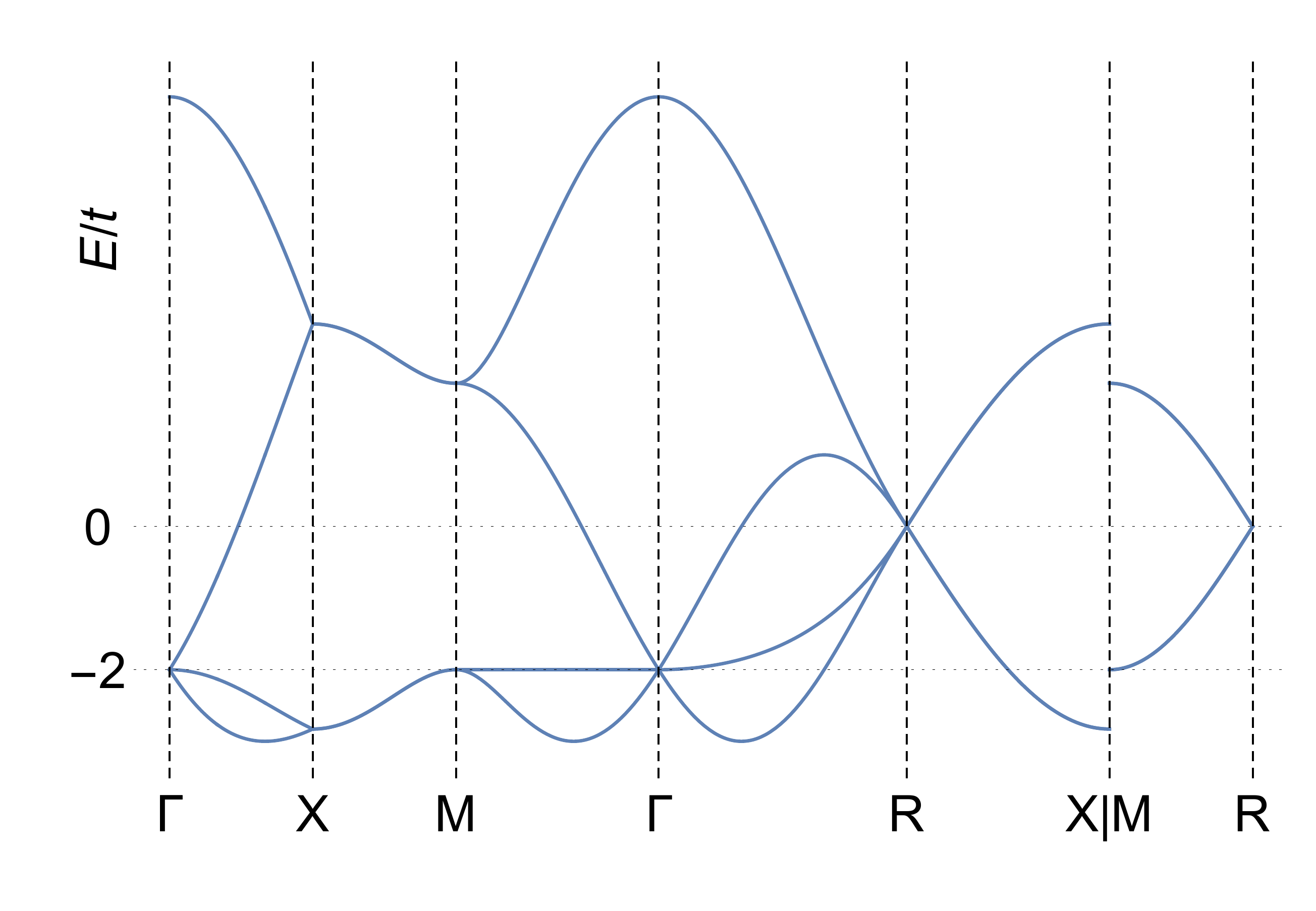}
     \caption{Band structure corresponding to Eq~(\ref{eq:Ham}). The threefold spin-$1$ fermion is visible at $\Gamma$ and the fourfold double spin-$1/2$ fermion at $R$. For a Fermi level between $-2t$ and $0$, the Fermi surfaces are similar to those sketched in Fig~\ref{fig:BZ}.}
     \label{fig:bands}
 \end{figure}

(The Hamiltonian for space group $P4_1 32$ is related by inversion.)
This Hamiltonian is equivalent to the one studied for B20 materials in the Supplementary Information of \cite{chang2017unconventional} with $v_1 = \pm v_p$ and vanishing spin-orbit coupling. The band structure is plotted in Fig~\ref{fig:bands}.
 
The spectrum of the Hamiltonian in Eq.~(\ref{eq:Ham}) exhibits a threefold spin-1 fermion at $\Gamma$, and a double simple Weyl fermion at $R$. Ignoring spin degeneracy, each node has Chern number $\pm 2$. Therefore, an external interface has two Fermi arcs. For a crystal in space group $P4_3 32$, the fermion at $\Gamma$ is left handed while the one at $R$ is right handed. For a crystal in space group $P4_1 32$, they have opposite chirality.

\begin{figure}
    \centering
    \includegraphics[scale=0.2]{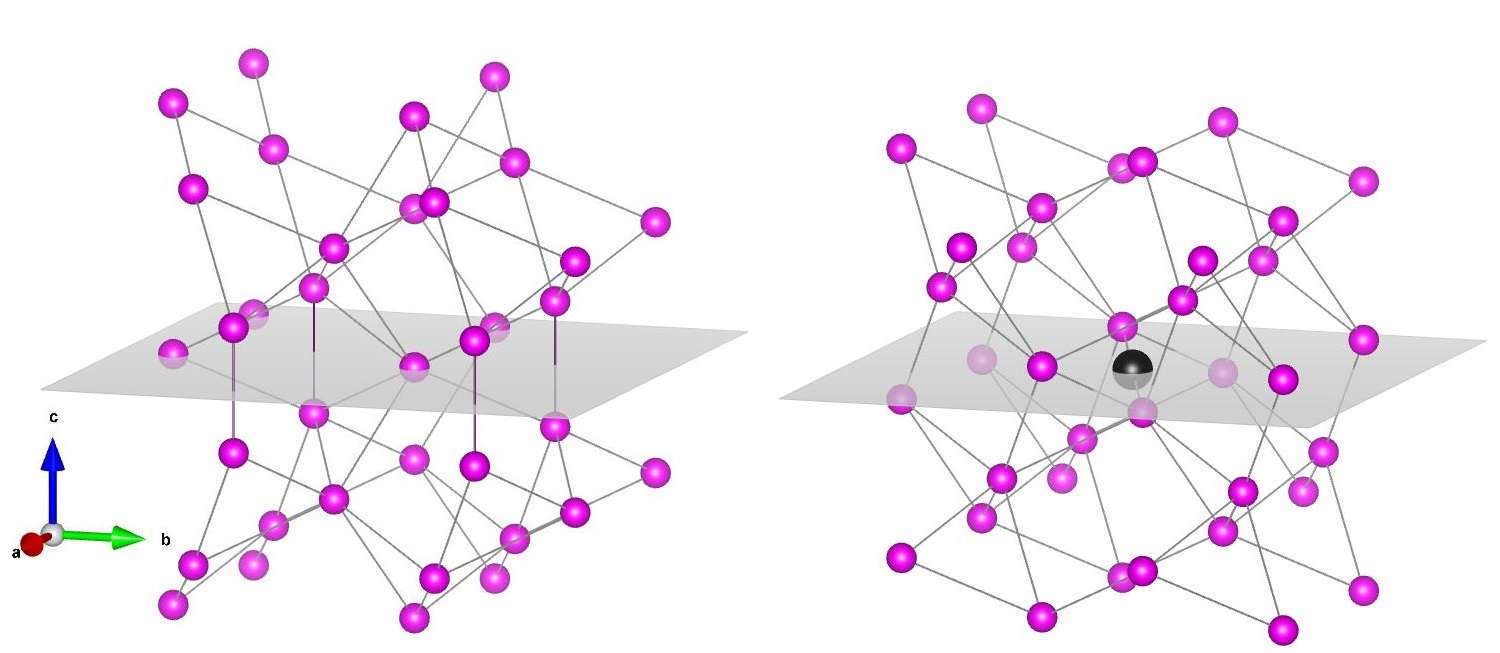}
    \caption{A mirror-twinned configuration (left), and an inversion-twinned configuration (right) of the toy model. The grey planes are the twin boundaries and the black dot is the inversion center. Figure created using the software VESTA \cite{VESTA}.}
    \label{fig:ToyModelTwin}
\end{figure}

We focus on merohedrally twinned crystals, with a twin boundary in a $(001)$ plane.
In the region above the boundary ($z>0$), the crystal is in the enantiomer in space group $P4_3 32$, 
while below the boundary ($z<0$), the crystal is in the space group $P4_1 32$. The chirality projection $\chi\hat{n}$ is in the direction $+\hat{z}$ at $\bar{\Gamma}$ and $-\hat{z}$ at $\bar{R}$. 
At least two kinds of twin boundaries are possible, corresponding to the configurations described in Sec.~\ref{sec:B20} and shown in Fig.~\ref{fig:B20twins}: 
the two enantiomers can be mapped to each other by either mirror symmetry or inversion symmetry across the twin boundary.
In the mirror-twinned system, there are atoms across the twinning plane that are separated by $a/2 $ instead of $\sim 0.612 a$; to construct the tight-binding model for the twinned crystal, we take the interaction between these pairs of atoms to have the same magnitude $t$ as the interactions between bulk nearest neighbors. The mirror-twinned and inversion-twinned boundaries are illustrated in Fig~\ref{fig:ToyModelTwin}.

\begin{figure}
    \centering
    \includegraphics[scale=0.081]{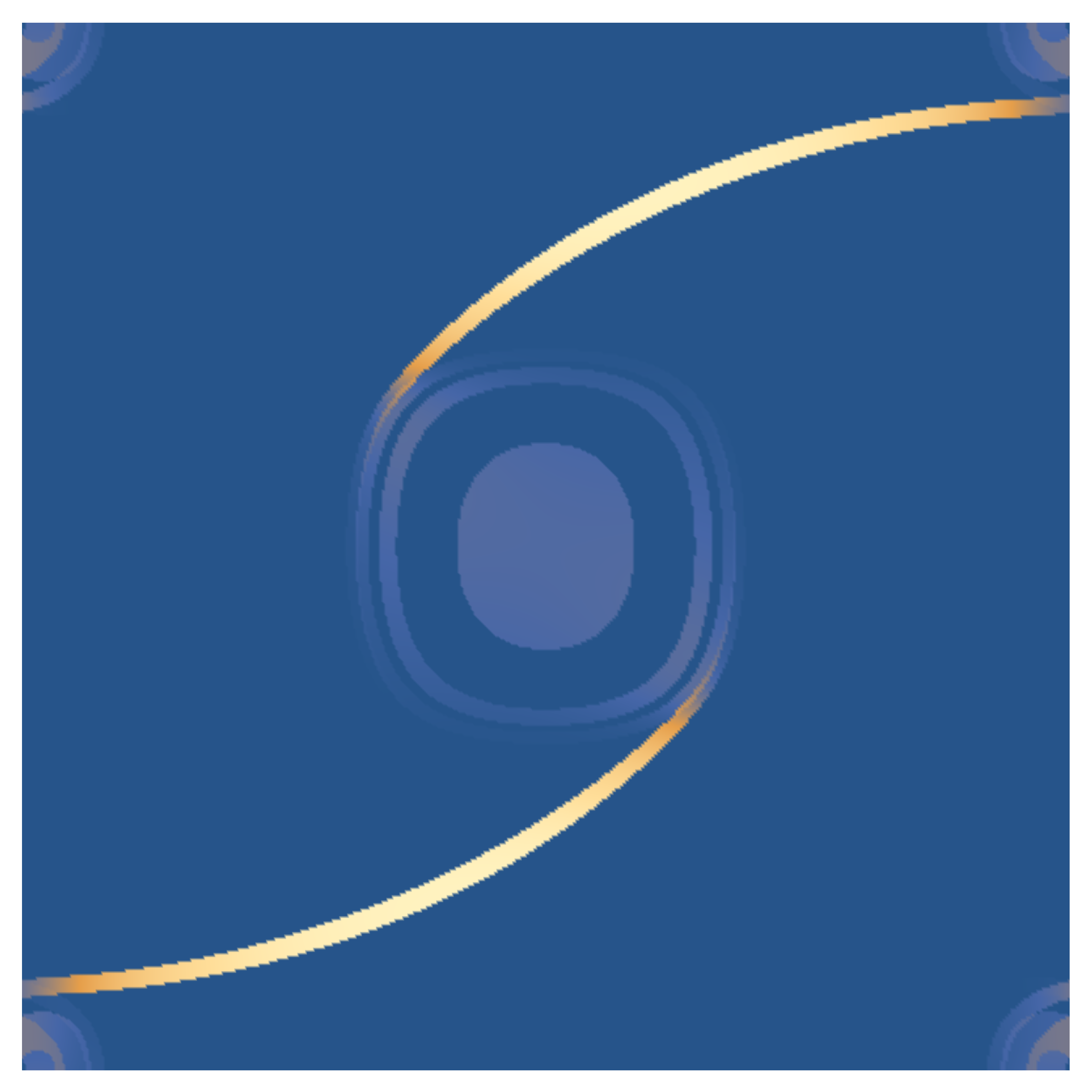}\hspace{0.02\textwidth}\includegraphics[scale=0.2]{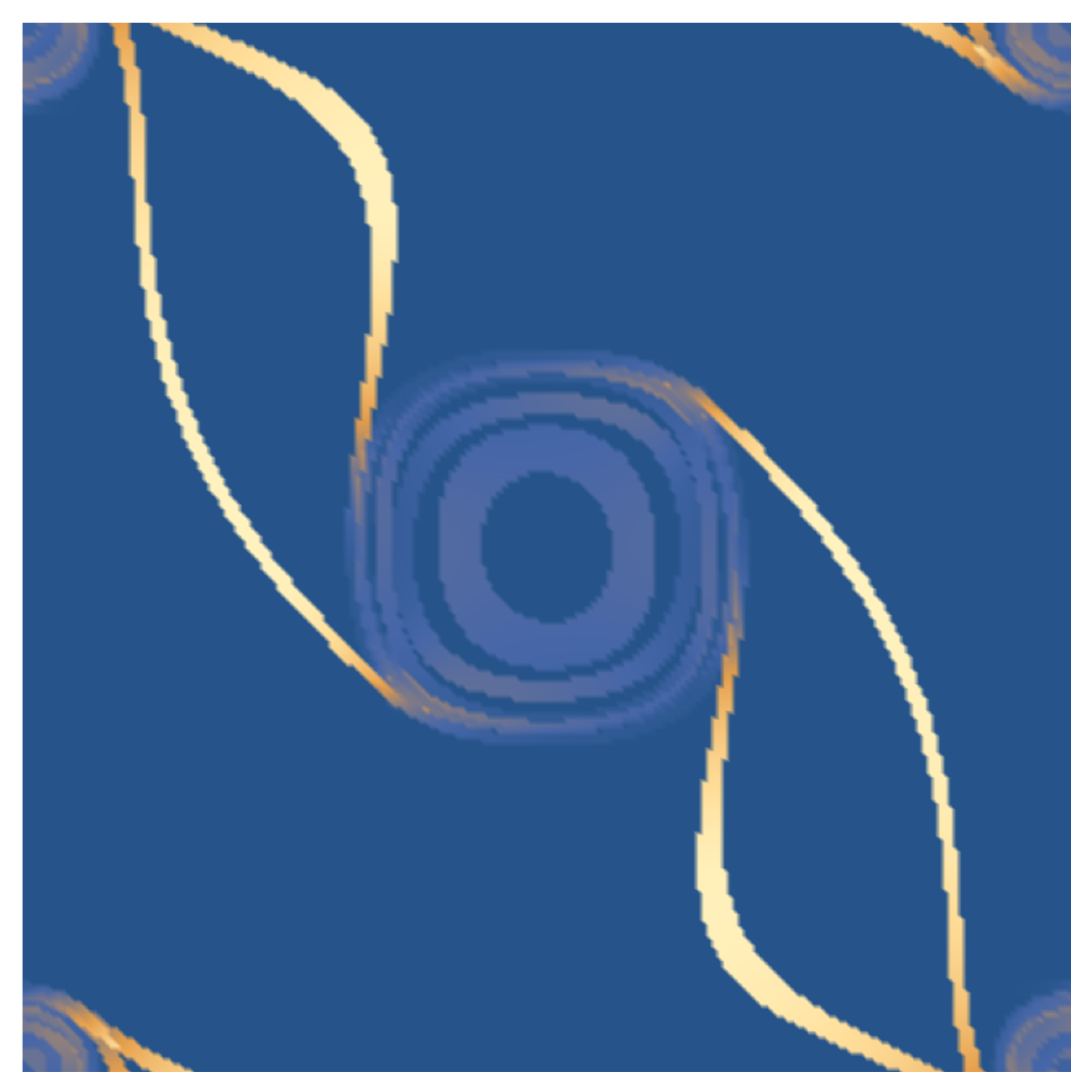}\hspace{0.02\textwidth}\includegraphics[scale=0.2]{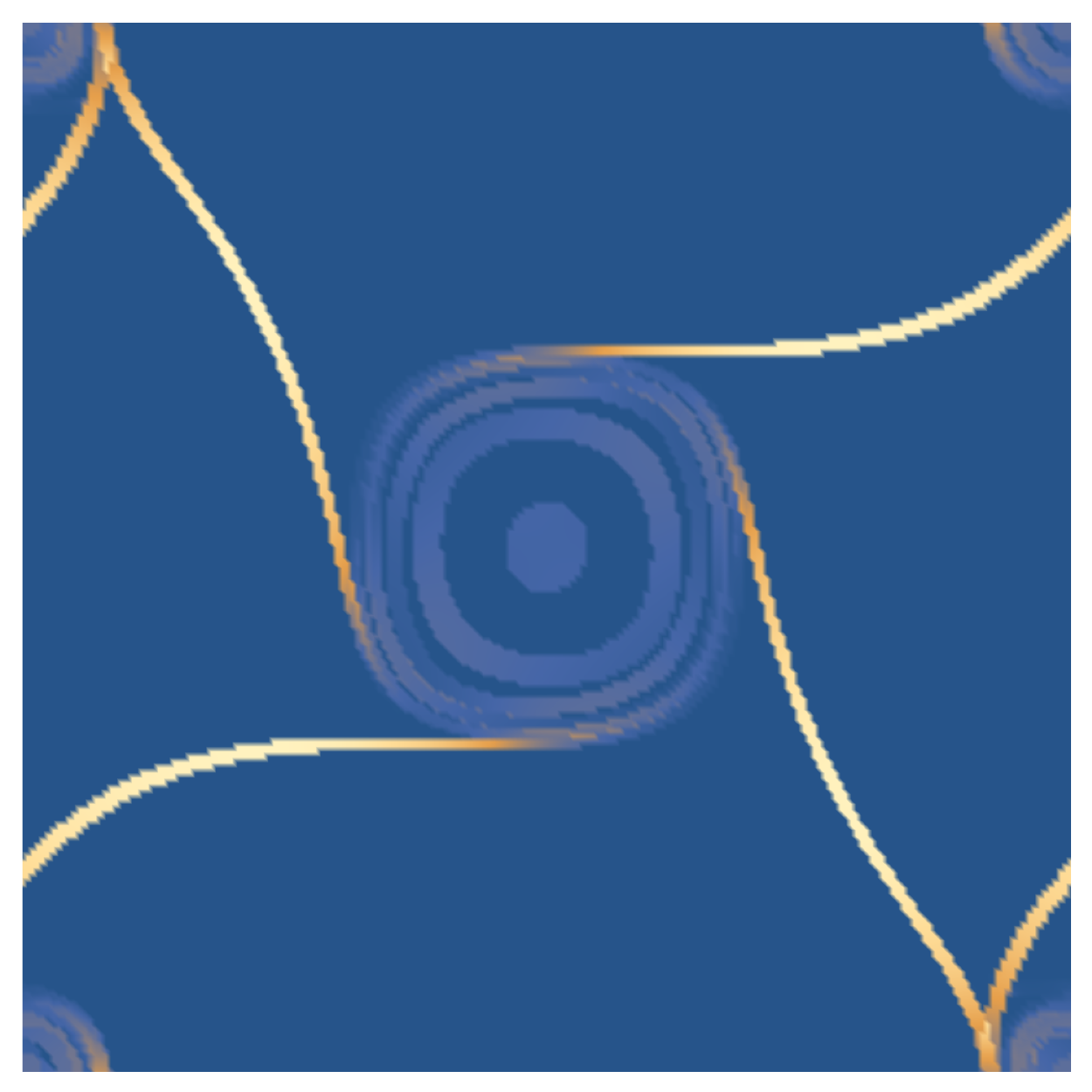}
    \caption{(L-R) Fermi arcs on an external boundary, a mirror-twinned boundary, and an inversion-twinned boundary for $E/t = -0.5\pm 0.05$.}
    \label{fig:EMIplot}
\end{figure}

We illustrate the Fermi arcs at an external boundary, a mirror-twinned boundary, and an inversion-twinned boundary in Fig~\ref{fig:EMIplot}: the connectivity can differ dramatically for the different types of boundaries. In these figures, the Fermi arcs are visualized by plotting the density of fermions with energies within $E\pm \delta E$ within one unit cell of the twin boundary. The bulk Fermi surfaces appear as diffuse regions, while the Fermi arcs appear as bright curves. 
In our finite slab, the projections of bulk states appear as discrete rings, but in a semi-infinite system, they would have a continuous spectrum. 
The Fermi arc states localized at the boundary are unaffected by the finite width of the slab.

The Fermi arcs are symmetric under $\vec{k}\to -\vec{k}$ due to time reversal symmetry. 
However, although the space groups $P4_3 32$ and $P4_1 32$ contain fourfold screw rotations, the Fermi arcs need not be (and are not) invariant under this symmetry, since it is not preserved on the surface. 
This asymmetry could result in anisotropic conductivity entirely due to the surface Fermi arcs, even though the bulk conductivity in a cubic crystal is isotropic.
For example, in the case of a mirror-twinned boundary, since the arcs are oriented along $(1\bar{1}0)$ and their velocity is along $(110)$, the conductivity would be enhanced along $(110)$ and suppressed along $(1\bar{1}0)$.

\begin{figure}
    \centering
    \includegraphics[scale=0.2]{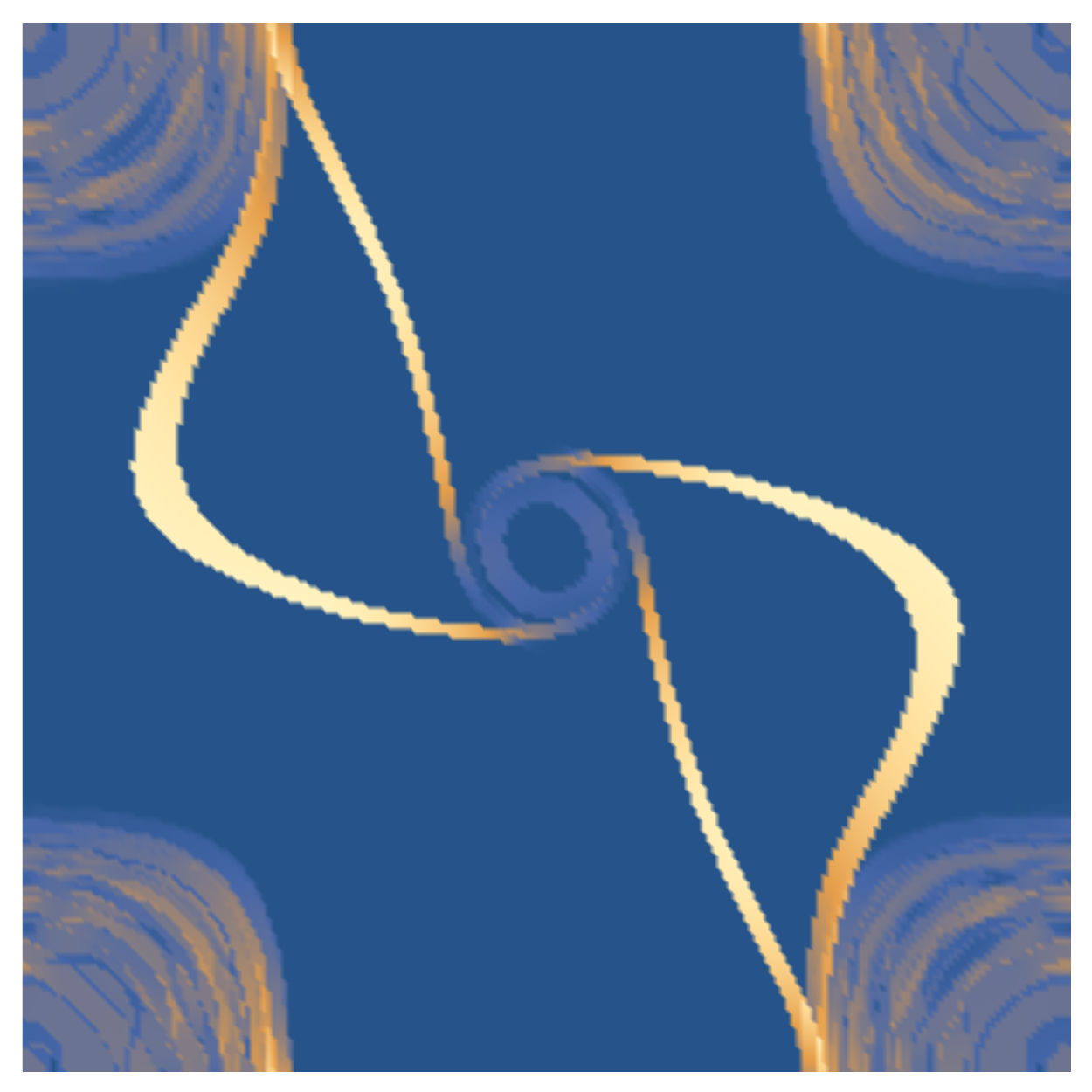}\hspace{0.02\textwidth}\includegraphics[scale=0.2]{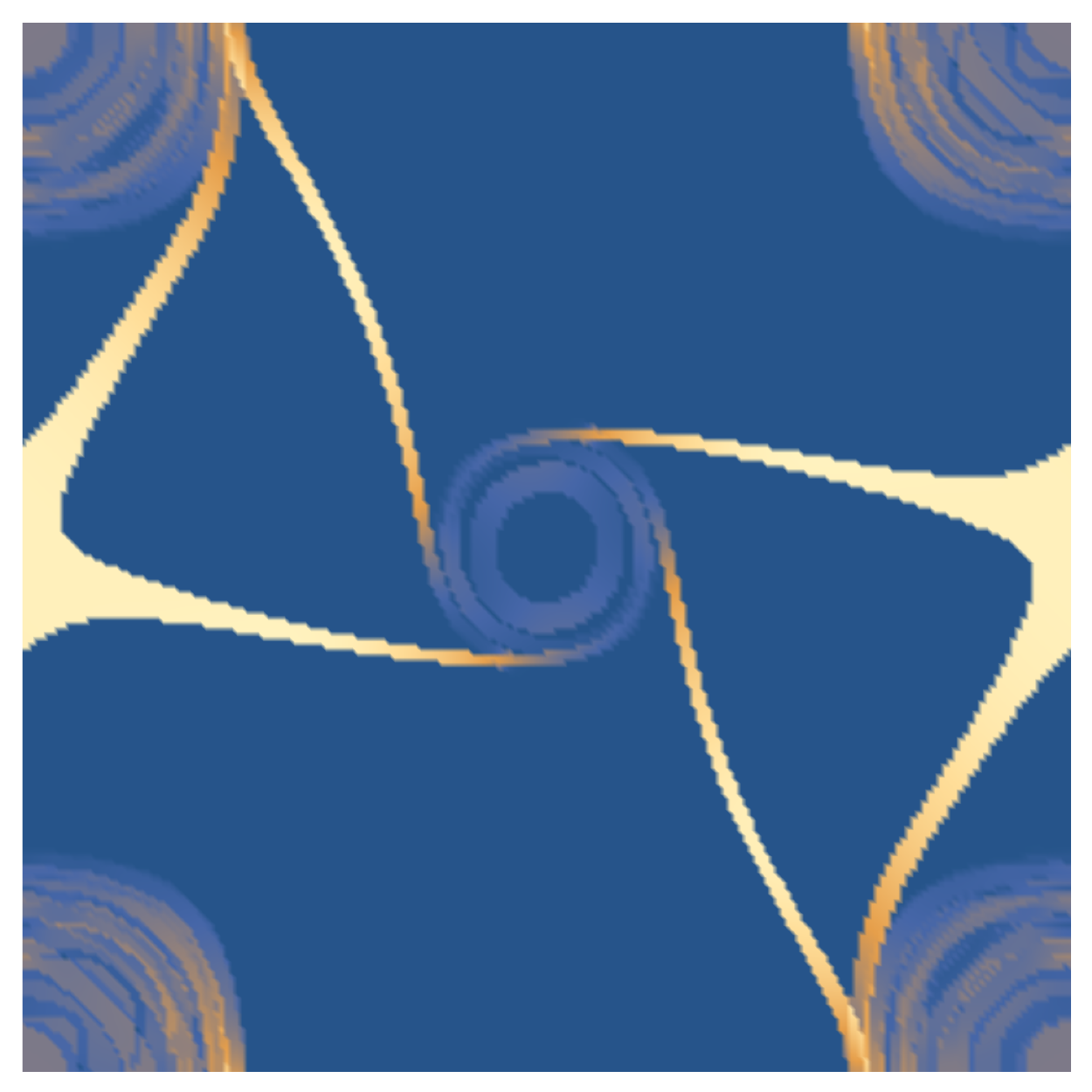}\hspace{0.02\textwidth}\includegraphics[scale=0.2]{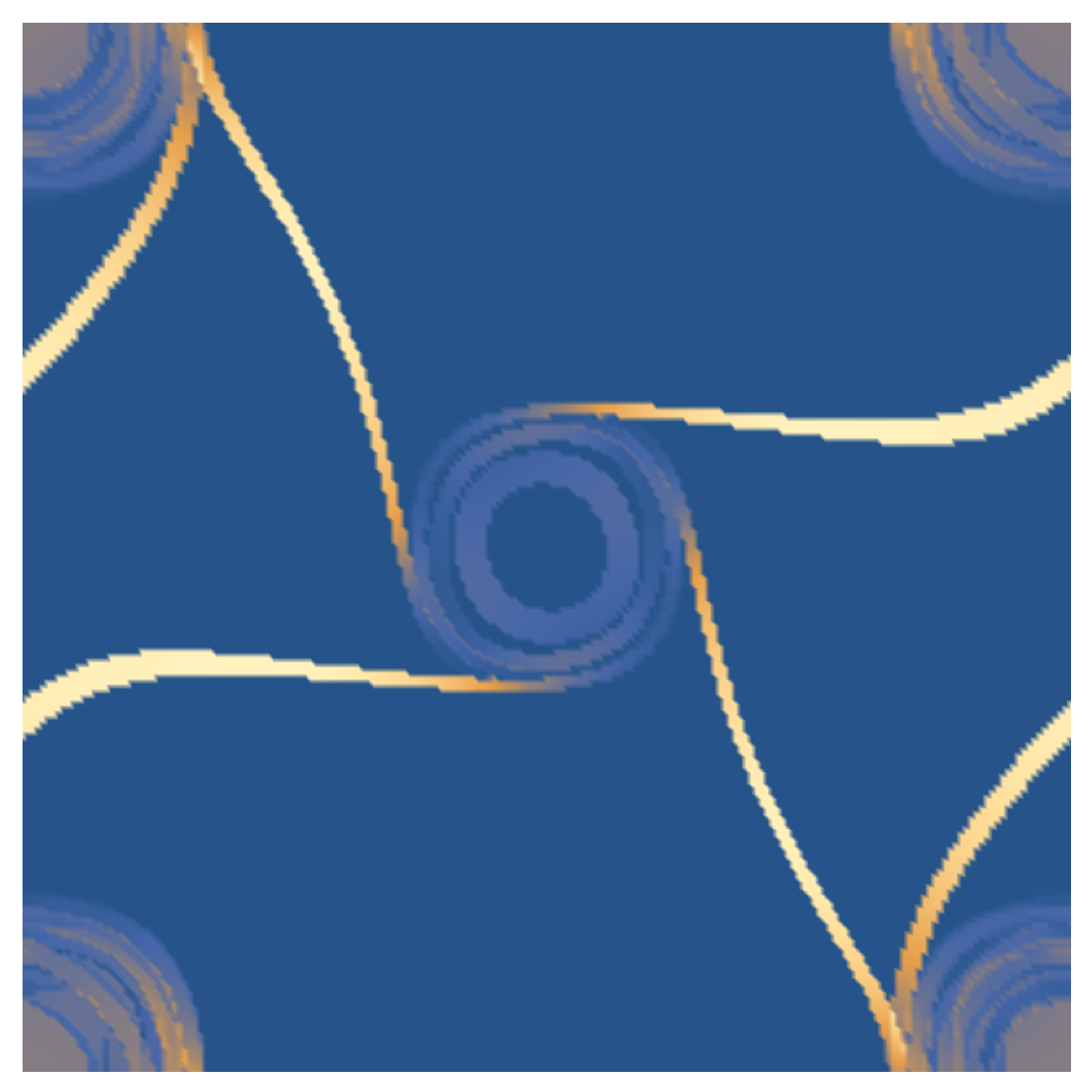}
    \caption{(L-R) Fermi arcs at an inversion-twinned boundary for $E/t = -1.4 \pm 0.05$, $-1.2\pm 0.05$, and $-1.0 \pm 0.05$, respectively.}
    \label{fig:energiesplot}
\end{figure}

The connectivity of the Fermi arcs can change with energy. We demonstrate this by plotting the arcs at an inversion-twinned boundary for different energies in Fig.~\ref{fig:energiesplot}. As the Fermi level changes, the arcs intersect and change their connectivity. This is similar to change in Fermi arc connectivity observed in doped RhSi and CoSi \cite{WeylCrossings}, where the \textit{external} arcs change connectivity, but here we are discussing internal Fermi arcs buried inside the material.

\begin{figure}
    \centering
    \includegraphics[scale=0.3]{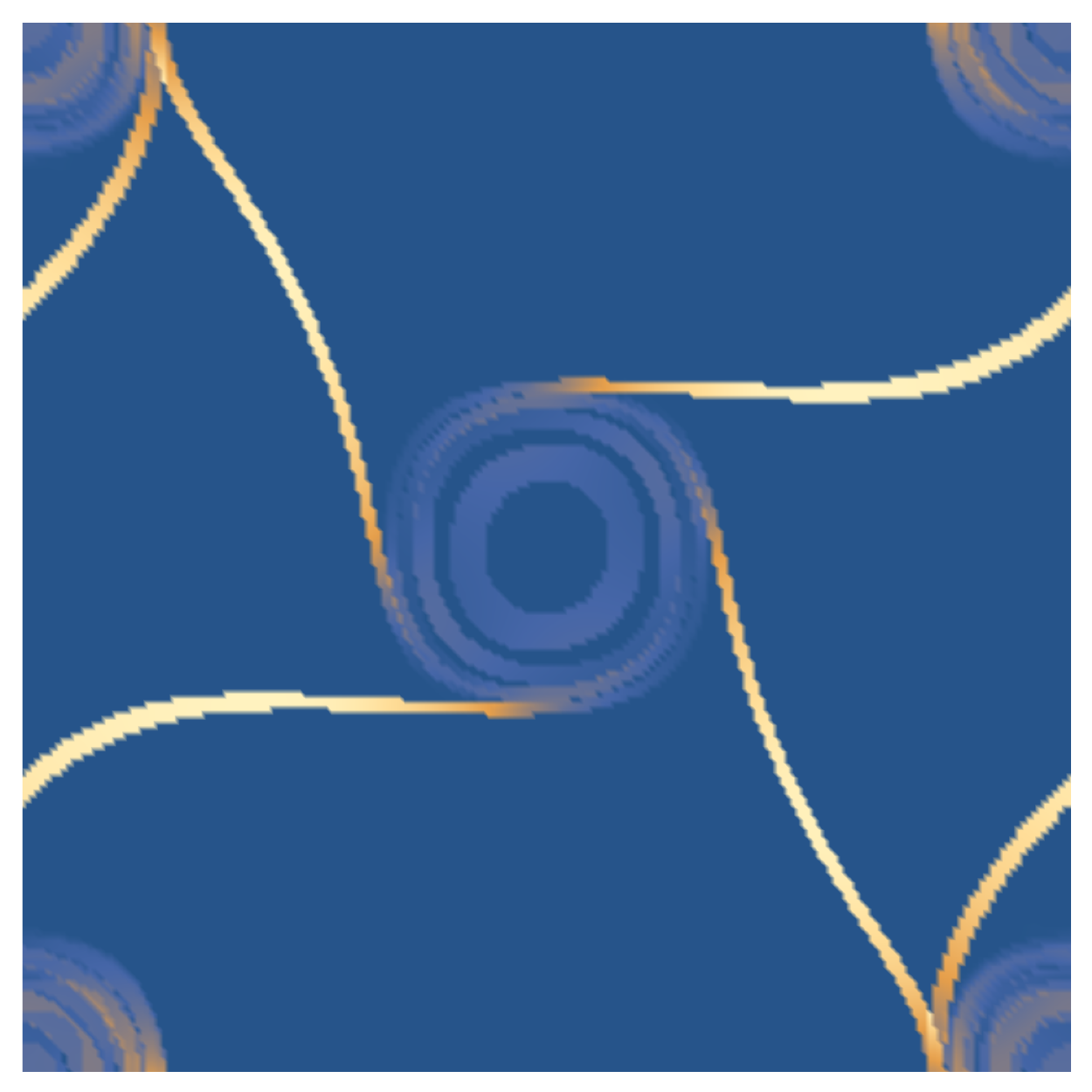}\hspace{0.01\textwidth}\includegraphics[scale=0.3]{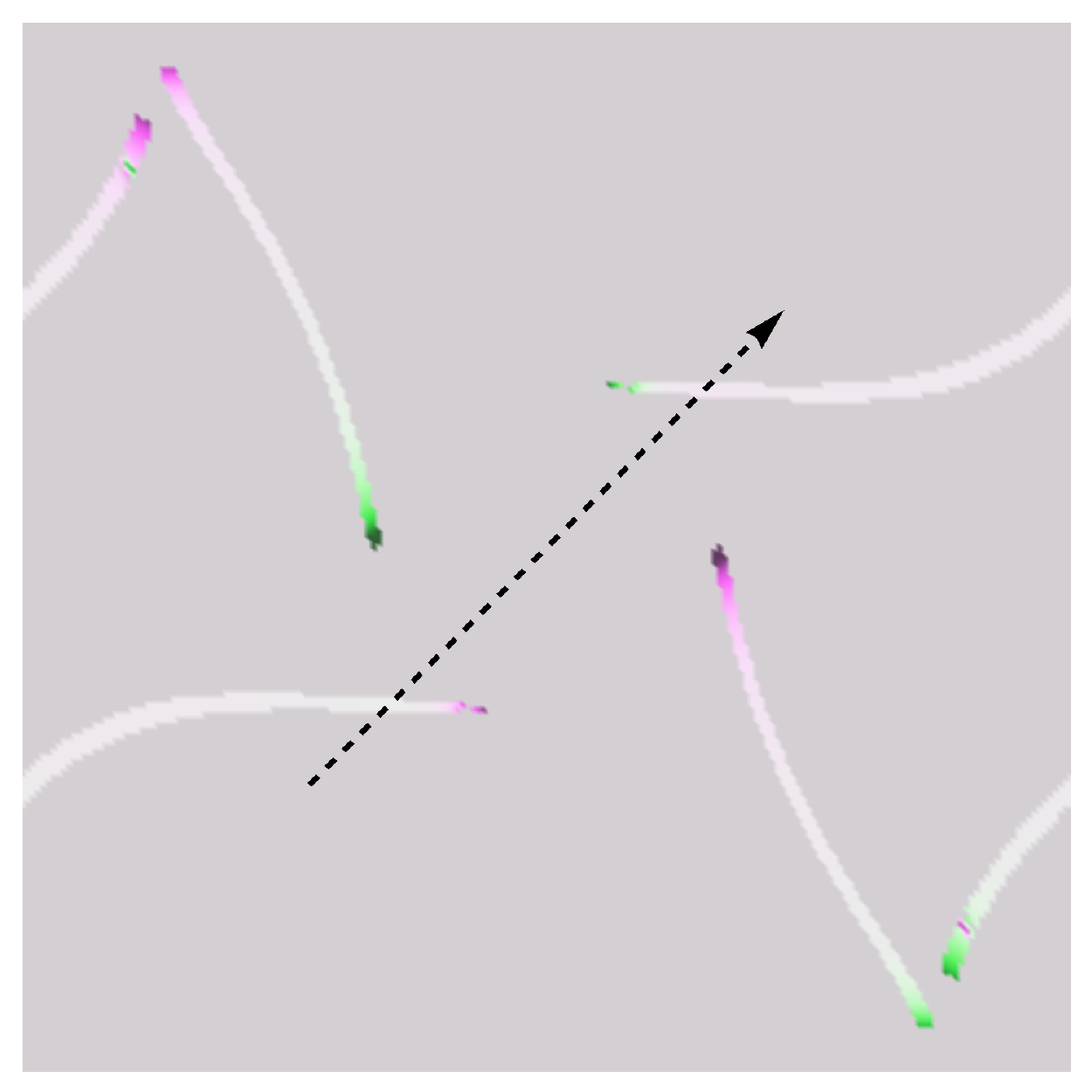}\includegraphics[scale=0.45]{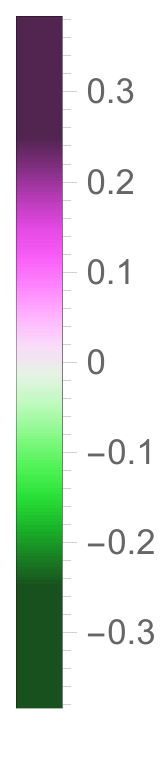}\\ \includegraphics[scale=0.3]{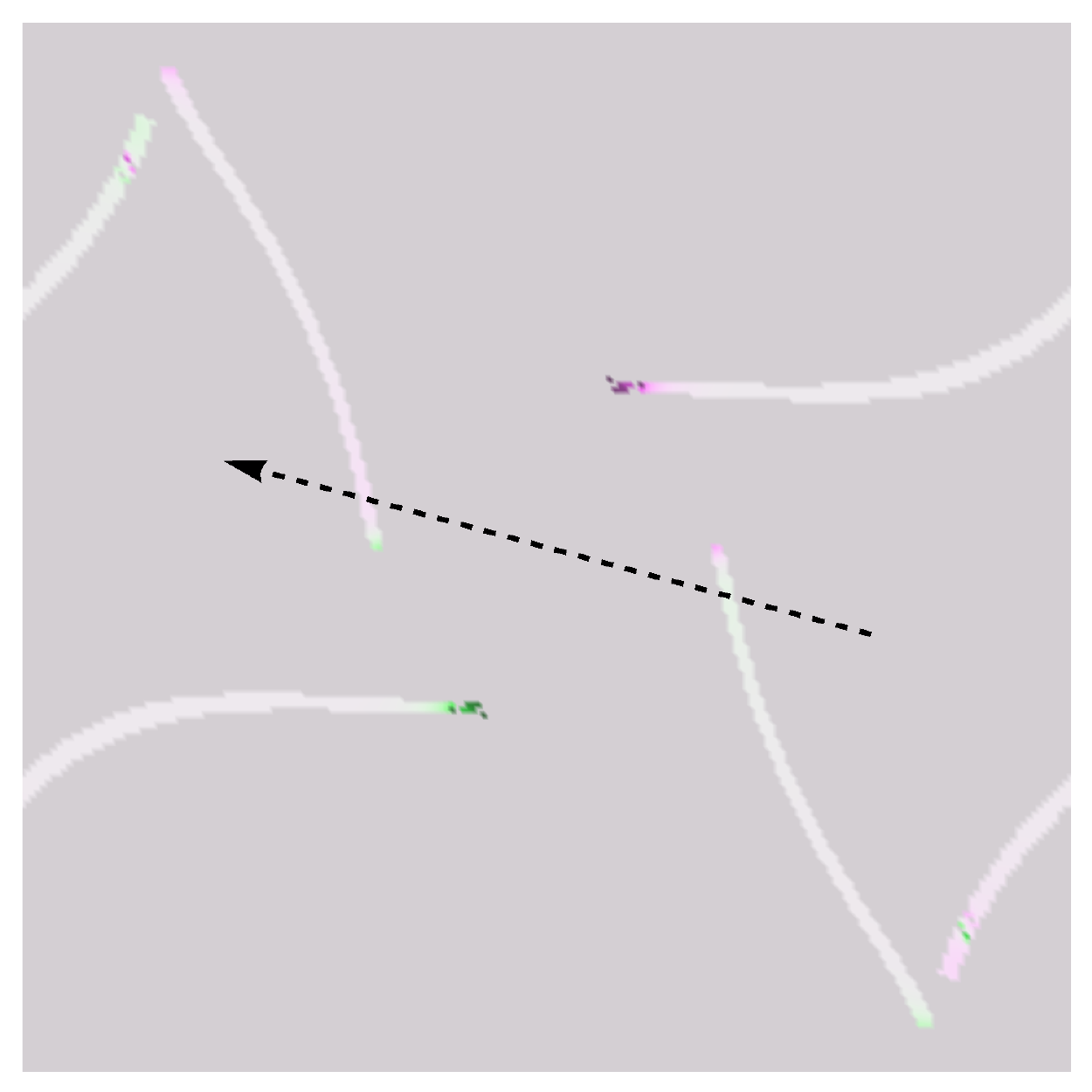}\hspace{0.01\textwidth}\includegraphics[scale=0.3]{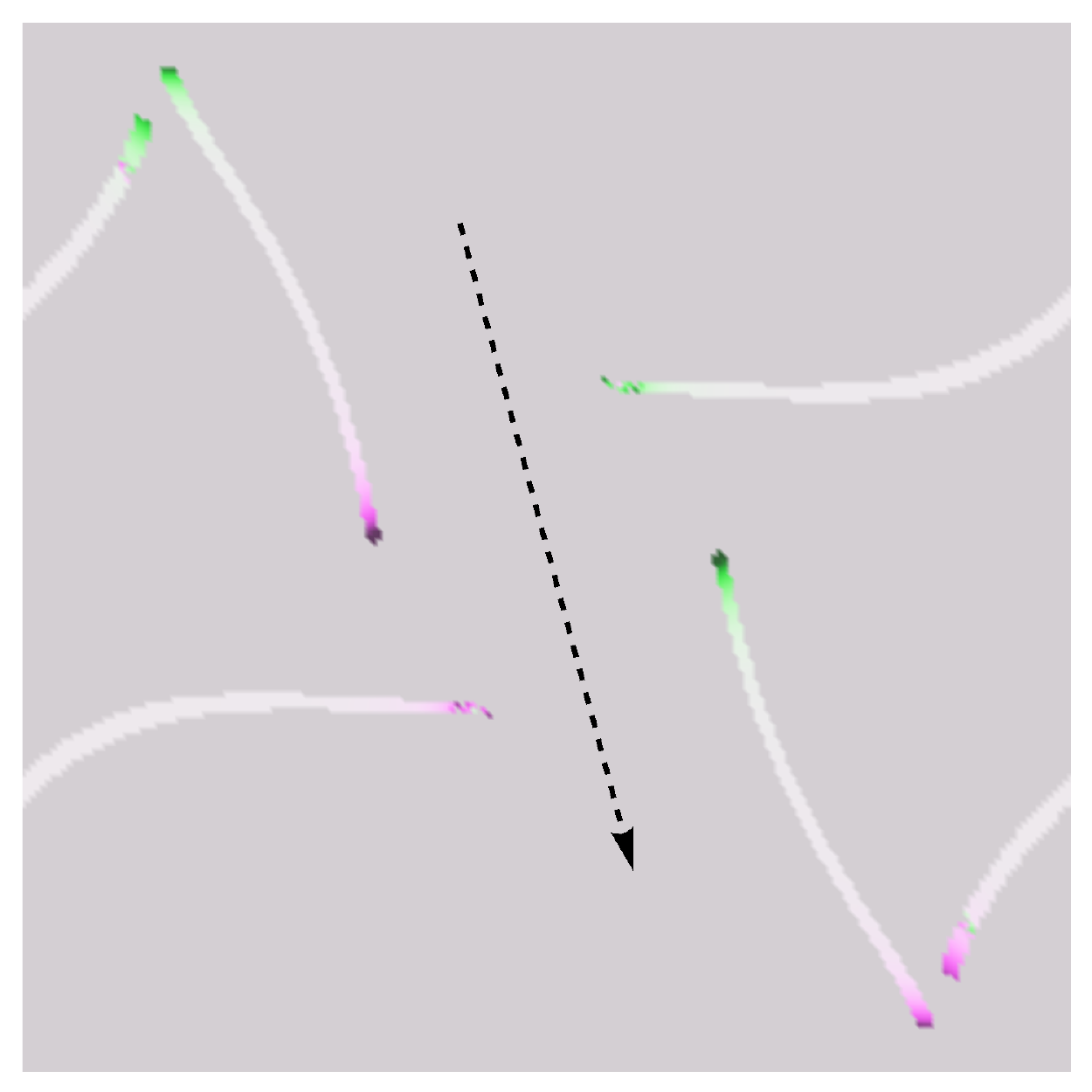}\includegraphics[scale=0.45]{MagArcsScale.pdf}
    \caption{Twin boundary Fermi arcs and their dehybridization with magnetic field for an inversion-twinned boundary with $E/t = -0.8\pm 0.05$. Color indicates $\langle z\rangle/a$; 
    since the Fermi arcs are localized to the boundary at $z=0$ except at their endpoints, only their ends appear in color.
    The dashed arrow indicates the direction of the magnetic field; its magnitude is $0.005a^{-2}\hbar/e$ for all plots.}
    \label{fig:magarcs}
\end{figure}

In the absence of magnetic field, all internal Fermi arcs are completely hybridized for both types of twin boundaries, i.e., the Fermi arcs cannot be associated with either side of the twin boundary and decay symmetrically into either side. However, because the ends of the Fermi arcs start to decay into the bulk, they can be dehybridized by an external in-plane magnetic field. We illustrate this in Fig~\ref{fig:magarcs} by plotting the expectation value of the position coordinate $\langle z\rangle$ perpendicular to the twin boundary for the internal Fermi arcs in an inversion-twinned system. The dehybridization of different arcs depends quite dramatically on the direction of the magnetic field. If the magnetic field is along $(110)$, all four arcs originate and terminate in opposite crystals. In this configuration, the arcs with positive $k_y$ correspond to right handed fermions, while the arcs with negative $k_y$ correspond to left handed fermions. From Eq~\ref{eq:veldir}, the velocity of the arc fermions is rotated by 90 degrees counterclockwise to the change in momentum, as measured from $\bar{\Gamma}$ to $\bar{M}$.
Therefore, if there is a positive chiral chemical potential $\mu_5 = (\mu_R-\mu_L)/2$, the quantized interface CME \textit{electron} current will be towards the left $(\bar{1}00)$; its sheet density will be $\frac{4e^2}{2\pi \hbar}\frac{1}{a}\mu_5$ (from Eq~\ref{eq:quantizedHall}). In contrast, the density of the interface Hall current (which is a major contribution to the magnetoconductivity) due to the bulk states \cite{CME} is $C\frac{e^3}{2\pi^2\hbar^2}\vec{B}\mu_5$, where $C=4$ is the Chern number.  If the lattice constant is $a = 0.5\mathrm{nm}$ and the magnetic field is $B = 13\ \mathrm{T} = 0.005a^{-2}\hbar/e$, the ratio of the bulk Hall current density to the twin boundary interface current density is $eBa/(\pi\hbar) \sim 3\times 10^6\  \mathrm{m}^{-1}$. 
Thus, if the thickness of the sample is of submicrometer scale, the effect of magnetic field on  transport will be dominated by the internal arcs.


\section{Internal Arcs in Cobalt Monosilicide}

To test the validity of our simplified tight binding model, we performed ab initio calculation within density functional theory (DFT) for the B20 material CoSi. We use the structural parameters as reported in Ref. \cite{TQC1,TQC2,TQC3,TQCServer}: the lattice constant is $a = 0.445\ \mathrm{nm}$ and the parameters $x$ and $y$ which appear in Table~\ref{tab:B20pos} are $0.1434$ and $0.1565$, corresponding to Co and Si, respectively.
We consider a twin boundary at $z=0.25a$ with an inversion center at $(0,0.4,0.25)a$ (which would exactly preserve nearest neighbor distances if $x$ and $y$ were both $0.15$).
The atoms are located at Wyckoff positions described by Table~\ref{tab:B20pos} for $z>0.25a$ and its inversion-image for $z<0.25a$. We consider a slab periodic in the $x$ and $y$ directions containing atoms at a distance up to $6.25a$ from the twin boundary. The slab contains $50$ atoms of each species per two-dimensional unit cell.
 
To study the electronic structure, we performed DFT calculations as implemented in the Vienna ab initio simulation package (VASP) \cite{VASP1,VASP2,VASP3,VASP4}. The interaction between ion cores and valence electrons was treated by the projector augmented-wave method \cite{DFT_PAW}, the generalized gradient approximation (GGA) for the exchange-correlation potential with the Perdew-Burke-Ernzerhof for solids parameterization \cite{DFT_PBE} and spin-orbit coupling was taken into account by the second variation method \cite{DFT_SOC}. The lattice parameters and atomic positions were fully optimized and obtained by minimization of the total energy of the bulk system. A Monkhorst-Pack centered at $\Gamma$ k-point grid of (11x11) for reciprocal space integration and 500 eV energy cutoff of the plane-wave expansion have been used for the self-consistent calculation. Fermi surface calculation was performed on a k-point grid of (21x21). The resulting calculation was analyzed in PyProcar \cite{DFT_PyProcar}.
 
 \begin{figure}
     \centering
     \includegraphics[width=\linewidth]{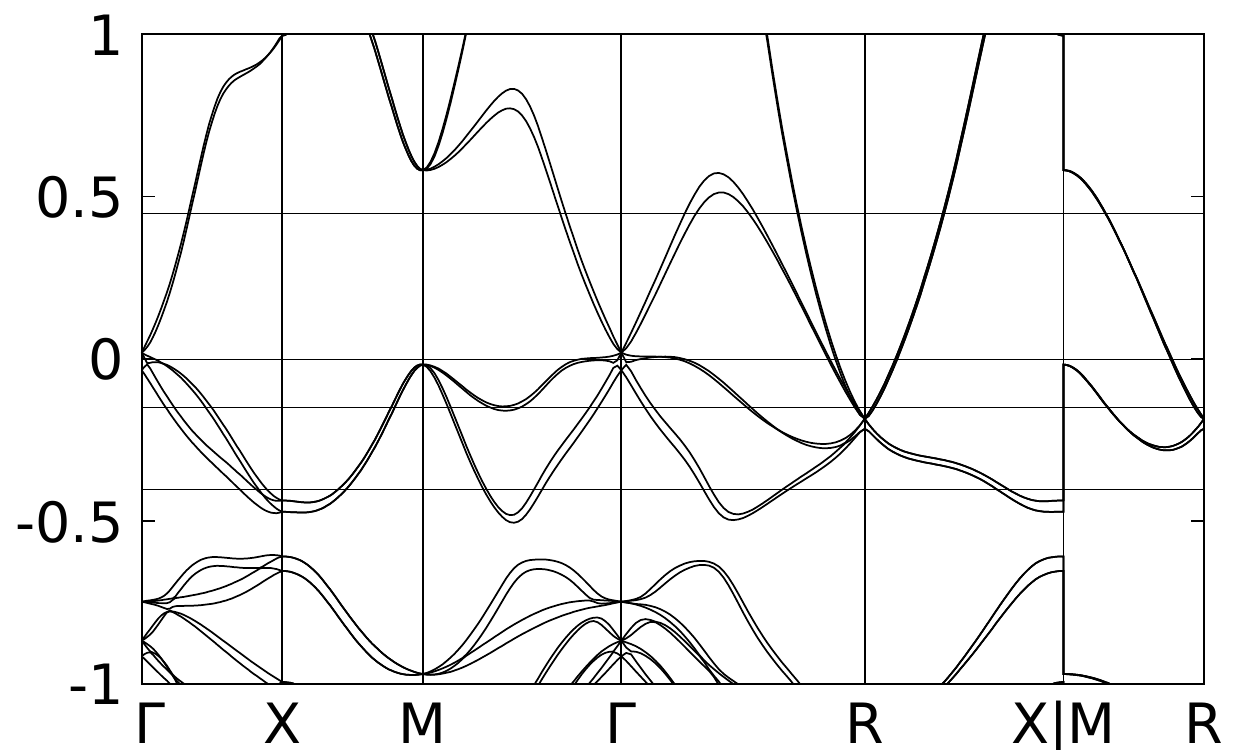}
     \caption{Electronic band dispersion of bulk CoSi. Solid horizantal lines denote Fermi surface calculation energies on Fig. \ref{fig:DFT_FS}.}
     \label{fig:DFT_Bands}
 \end{figure}
 

 \begin{figure}[t]
     \centering
     \includegraphics[width=\linewidth]{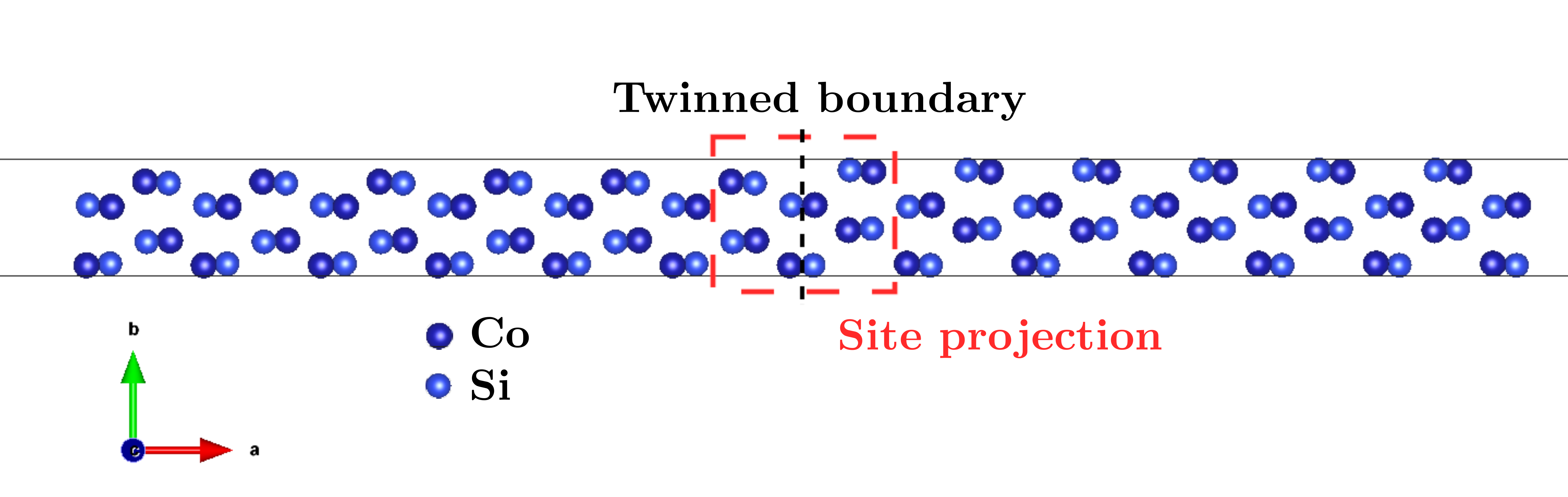}
     \includegraphics[scale=0.3,trim={2cm 0 4.8cm 0},clip]{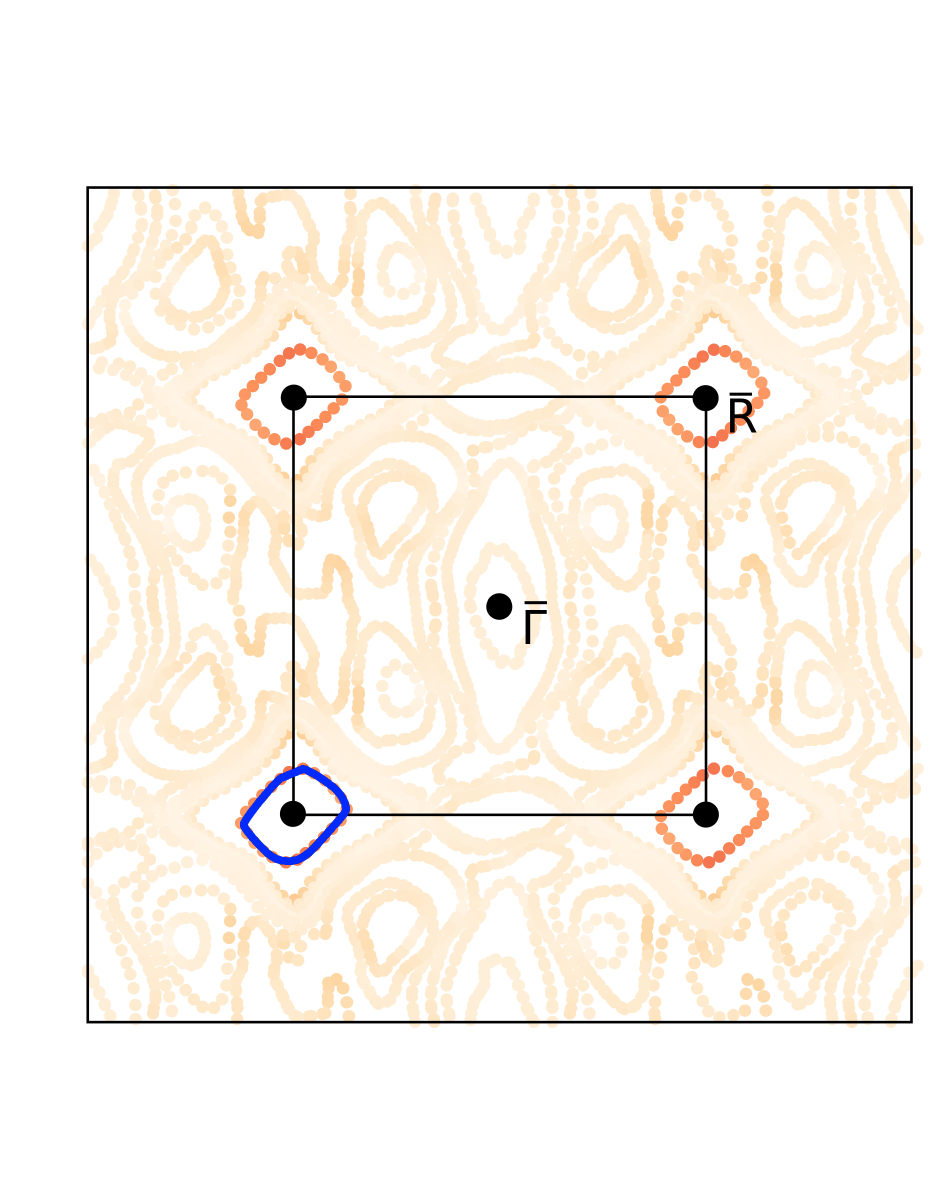}\includegraphics[scale=0.3,trim={2cm 0 4.8cm 0},clip]{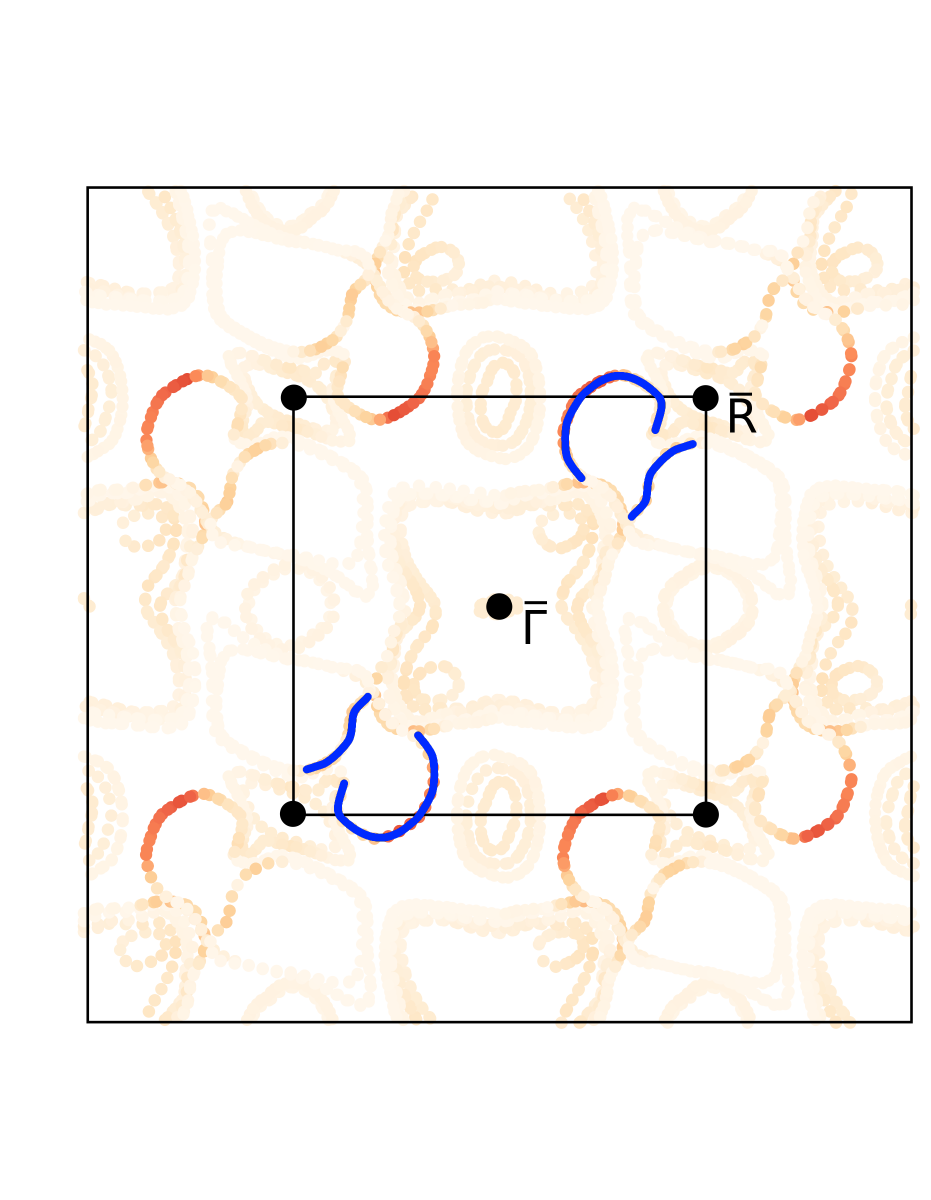}\includegraphics[scale=0.3,trim={2cm 0 3cm 0},clip]{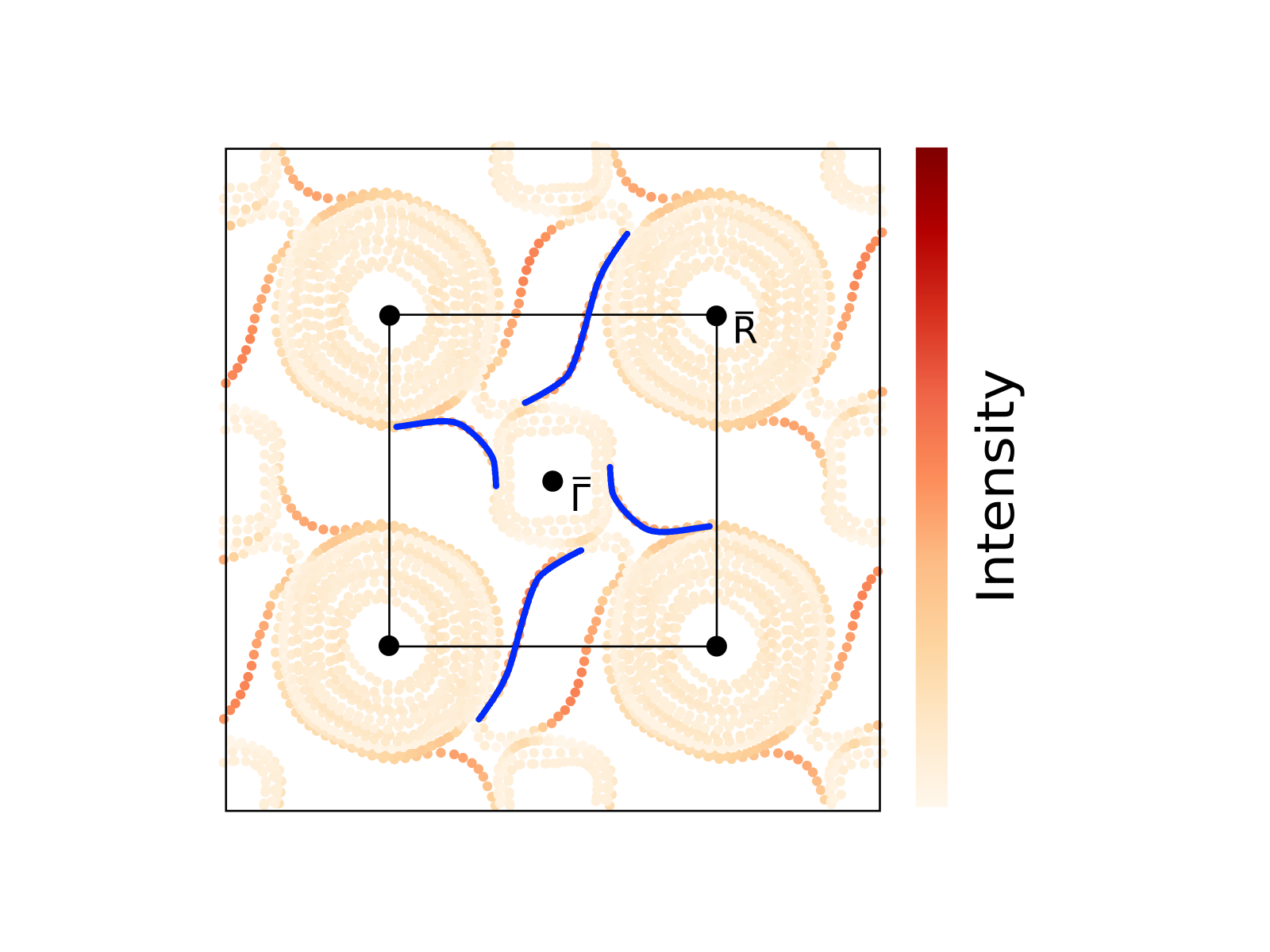}
     \caption{Top: inversion-twinned slab used in the ab initio calculation. The black line indicates the twin boundary.  Bottom: Fermi surface calculation projected onto the red box shown surrounding the twin boundary shown in the top figure,
     for $E-E_F=-0.40\text{eV}$, $E-E_F=-0.15\text{eV}$ and $E-E_F=0.45\text{eV}$, respectively, with $E_F$ the Fermi energy. The black square indicates the Brillouin zone at the twin boundary. The blue lines indicate states localized to the twin boundary.}
     \label{fig:DFT_FS}
 \end{figure}
 
 
Fig. \ref{fig:DFT_Bands} shows the bulk band dispersion of CoSi, which displays qualitative agreement with the tight binding model dispersion plotted in \ref{fig:bands}. One effect of spin-orbit coupling is that the fermion at $\Gamma$ splits into a fourfold and a twofold fermion, while the one at $R$ splits into a double threefold and a doubly degenerate trivial fermion \cite{CanoMultifold,FlickerMultifold}. The splitting at $\Gamma$ is $54$ meV and the splitting at R is $31$ meV. Since the splitting is significantly smaller than the other energy scales in this system, we can ignore the spin-orbit coupling.

Fig. \ref{fig:DFT_FS} shows the interface states for three representative energies: $E-E_F=-0.40\text{eV}$, $-0.15\text{eV}$, and $0.45\text{eV}$, where $E_F$ is the Fermi energy. The color scale indicates the projection on the region, defined in Fig. \ref{fig:DFT_FS} inside the red dashed box. In this picture, the Fermi arcs and other surface states appear as isolated curves, while bulk states appear as clusters of curves close to each other. The bulk states have a projection of $\sim 0.12$ on average (because the region has 6 out of 50 atoms). The internal Fermi arcs can be distinguished from the external Fermi arcs and the bulk states as they have a projection significantly greater than $0.12$. The bright red curves are the internal Fermi arcs, while the fainter yellow curves are the bulk states and the external Fermi arcs. These plots cover multiple Brillouin zones; the internal arcs in one Brillouin zone have been highlighted in blue.
These three Fermi energies exhibit distinct behaviors, as we now describe.

At $E-E_F=-0.40\ \text{eV}$, there are no disjoint Fermi surfaces, but there are pockets with localized interface states that form closed loops. Because these states form a closed Fermi interface, they would contribute to quantum oscillations; their frequency, determined by the area of the loop, would be $\sim 750\ \mathrm{T}$.

At $E-E_F=-0.15\ \text{eV}$, there are disjoint Fermi surfaces at $\Gamma$ and $R$; they are connected by four Fermi arcs, which form two loops. Therefore, the conductivity across the interface would exhibit quantum oscillations, whose frequency would also be equal to the area of the loop, $\sim 1200\ \mathrm{T}$.
This is similar to the Fermi arcs in our tight-binding model on the left side of Fig~\ref{fig:energiesplot}. 
At $E-E_F=0.45\ \text{eV}$, there are still disjoint Fermi surfaces, connected by four arcs, but they do not form closed loops, instead radiating in four different directions, in a pattern similar to the right side of Fig~\ref{fig:energiesplot}.

Our explicit demonstration of internal Fermi arcs in CoSi in Fig.~\ref{fig:DFT_FS} confirms that internal Fermi arcs will be present in real materials.
Further, we have shown that not only can the connectivity of internal Fermi arcs change with energy, but even their topology can change, i.e., Fermi arcs can change from forming a closed loop to radiating in four different directions; this feature was also observed in our tight-binding model in Fig.~\ref{fig:energiesplot}.
If the Fermi level can be brought into the regime where the Fermi arcs exhibit closed loops at the twin boundary, then transport measurements would exhibit quantum oscillations, as described in Sec.~\ref{sec:QO}.
The Fermi arcs may also contribute to the quantized boundary chiral mangetic current, as described in Sec.~\ref{sec:AHE}.



\section{Discussion}

Interfaces between Weyl materials with opposite chirality, such as twin boundaries and domain walls, can host interface-localized Fermi arcs, which are subject to topological constraints similar to those that govern surface Fermi arcs.
Such internal Fermi arcs have been predicted \cite{schroter2020observation,TwistArcs1,TwistArcs2,MagWallArcs1,MagWallArcs2,MagWallArcs3} but not observed. In this work, we propose two experiments to observe these Fermi arcs. 
First, if the internal arcs form a simply connected loop, they contribute to quantum oscillations with a frequency that depends only on the component of the magnetic field perpendicular to the interface. The frequency of the quantum oscillations is given by the area in momentum space spanned by the arcs. 
Second, we proposed a quantized sheet chiral mangetic effect effect due to dehybridization of Fermi arcs into either side of the boundary in an applied magnetic field.

To show the existence and estimate the size of internal Fermi arcs, we considered a twin boundary in a simplified and ab initio model of a B20 crystal.
We showed that the connectivity of internal arcs depends on the energetic details of the interface, i.e., the internal arcs are not identical to a ``sum'' of external arcs for both crystals. 
We also demonstrated that an external magnetic field dehybridizes the ends of the Fermi arcs. 
Our results provide evidence that the internal Fermi arcs localized on the twin boundary could be observed by the experiments we proposed.



Our theoretical investigation of topological
features originating from internal Fermi arcs encourages
development of techniques for the synthesis and measurement of chiral
twinned Weyl semimetals.
It will also spur an investigation into other characterization techniques, such as scanning tunneling microscopy and optical and acoustic probes to study the intriguing physics of these arcs.


\begin{acknowledgments}
S.K. thanks Claudia Felser, Egor Babaev, Evan Philip, Diptarka Hait, and Pavlo Sukhachov for helpful discussions.
N.M. acknowledges helpful discussion with Fang Yuan.

Nordita is supported in part by NordForsk.
J.C. acknowledges the support of the Flatiron Institute, a division of Simons Foundation.
This material is based upon work supported in part by the US Department of Energy under Award DE-SC0017662 (S.K.), the US National Science Foundation under Grant No. DMR-1942447 (J.C.), and the Gordon and Betty Moore Foundation’s EPIQS initiative through Grant GBMF9064 (N.M.). M.G.V.  and I.R. acknowledge the Spanish Ministerio de Ciencia e Innovaci\'{o}n (grant PID2019-109905GB-C21), Programa Red Guipuzcoana de Ciencia, Tecnolog\'{i}a e Innovaci\'{o}n 2021 No. 2021-CIEN-000070-01 Gipuzkoa Next and the Deutsche Forschungsgemeinschaft (DFG, German Research Foundation) GA 3314/1-1 – FOR 5249 (QUAST).
L.M.S. acknowledges support by the Gordon and Betty Moore foundation (EPIQS initiative), award number GBMF9064, by the Princeton Center for Complex Materials, a US National Science Foundation (NSF)-MRSEC program (DMR-2011750), the Packard and Sloan foundation.
\end{acknowledgments}

\bibliographystyle{apsrev4-1}
\bibliography{refs}
\end{document}